\documentclass[rmp,aps,nofootinbib,twocolumn,superscriptaddress]{revtex4}
\pdfoutput=1

\usepackage{graphics}
\usepackage{epsfig}

\usepackage{amsmath}
\usepackage{amssymb}
\usepackage{graphicx}% Include figure files
\usepackage{epsfig}
\usepackage{amsmath}
\usepackage{amssymb}
\usepackage{textcomp}
\usepackage{dcolumn}% Align table columns on decimal point
\usepackage{bm}% bold math

\newcommand{\ket}[1]{\ensuremath{\lvert #1 \rangle}}
\newcommand{\bra}[1]{\ensuremath{\langle #1 \rvert}}
\newcommand{\bracket}[2]{\ensuremath{\langle #1 \vert #2 \rangle}}

\newcommand{\ud}{\,\mathrm{d}}
\renewcommand{\vec}[1]{\ensuremath{\mathbf{#1}}}
\newcommand{\psig}{\phi}

\def\inbar{\,\vrule height1.5ex width.4pt depth0pt}
\def\IR{\relax{\rm I\kern-.18em R}}
\def\IC{\relax\hbox{$\inbar\kern-.3em{\rm C}$}}

\begin{document}
\title{Colloquium: Trapped ions as quantum bits -- essential numerical tools}

\author{Kilian Singer}
\email{email@kilian-singer.de}
\affiliation{Institut f\"ur Physik, Johannes Gutenberg-Universit\"at Mainz, 55099 Mainz, Germany}
\affiliation{Institut f\"ur Quanteninformationsverarbeitung,
Universit\"at Ulm, Albert-Einstein-Allee 11, 89069 Ulm, Germany}
\author{Ulrich Poschinger}
\affiliation{Institut f\"ur Physik, Johannes Gutenberg-Universit\"at Mainz, 55099 Mainz, Germany}
\affiliation{Institut f\"ur Quanteninformationsverarbeitung,
Universit\"at Ulm, Albert-Einstein-Allee 11, 89069 Ulm, Germany}
\author{Michael Murphy}
\affiliation{Institut f\"ur Quanteninformationsverarbeitung,
Universit\"at Ulm, Albert-Einstein-Allee 11, 89069 Ulm, Germany}
\author{Peter A. Ivanov}
\affiliation{Institut f\"ur Physik, Johannes Gutenberg-Universit\"at Mainz, 55099 Mainz, Germany}
\affiliation{Institut f\"ur Quanteninformationsverarbeitung,
Universit\"at Ulm, Albert-Einstein-Allee 11, 89069 Ulm, Germany}
\author{Frank Ziesel}
\affiliation{Institut f\"ur Physik, Johannes Gutenberg-Universit\"at Mainz, 55099 Mainz, Germany}
\affiliation{Institut f\"ur Quanteninformationsverarbeitung,
Universit\"at Ulm, Albert-Einstein-Allee 11, 89069 Ulm, Germany}
\author{Tommaso Calarco}
\affiliation{Institut f\"ur Quanteninformationsverarbeitung,
Universit\"at Ulm, Albert-Einstein-Allee 11, 89069 Ulm, Germany}
\author{Ferdinand Schmidt-Kaler}
\affiliation{Institut f\"ur Physik, Johannes Gutenberg-Universit\"at Mainz, 55099 Mainz, Germany}
\affiliation{Institut f\"ur Quanteninformationsverarbeitung,
Universit\"at Ulm, Albert-Einstein-Allee 11, 89069 Ulm, Germany}

\begin{abstract}
 Trapped, laser-cooled atoms and ions are quantum systems which can be experimentally controlled with an as yet unmatched degree of precision.
 Due to the control of the motion and the internal degrees of
  freedom, these quantum systems can be adequately described by a well known Hamiltonian. In this colloquium, we present powerful numerical tools for the optimization of the external control of the motional and internal states of trapped neutral atoms, explicitly applied to the case of trapped laser-cooled ions in a segmented ion-trap. We then delve into
  solving inverse problems, when optimizing trapping potentials for ions. Our presentation is
  complemented by a quantum mechanical treatment of the wavepacket
  dynamics of a trapped ion. Efficient numerical solvers
  for both time-independent and time-dependent problems are provided.
  Shaping the motional wavefunctions and optimizing a quantum gate is
  realized by the application of quantum optimal control
  techniques. The numerical methods presented can also be
  used to gain an intuitive understanding of quantum experiments with
  trapped ions by performing virtual simulated experiments on a personal
  computer. Code and executables are
  supplied as supplementary online material\footnote{Download source
    code and script packages (no compiler needed) optimized for Linux and Windows at
    \texttt{http://kilian-singer.de/ent}.}.
\end{abstract}

%\date{May 2001}
\maketitle
\tableofcontents

\section{Introduction}
\label{sec:intro} The information carrier used in computers is a
bit, representing a binary state of either zero or one. In the
quantum world a two-level system can be in any superposition of
the ground and the excited state. The basic information carrier encoded by such a
system is called a quantum bit (qubit). Qubits are manipulated by quantum gates---unitary transformations
on single qubits and on pairs of qubits.

Trapped ions are among the most promising physical systems for
implementing quantum computation \citep{NC}. Long coherence times
and individual addressing allow for the experimental
implementation of quantum gates and quantum computing protocols
such as the Deutsch-Josza algorithm \citep{Gulde:2003},
teleportation \citep{Riebe:2004,Barrett:2004}, quantum error
correction \citep{Chiaverini:2004}, quantum Fourier transform
\citep{Chiaverini:2005} and Grover's search \citep{Brickman:2005}.
Complementary research is using trapped neutral atoms in micro
potentials such as magnetic micro traps
\citep{Fortagh:2007,Schmiedmayer:2002}, dipole traps
\citep{Grimm:2000,Gaetan:2009} or optical lattices
\citep{Lukin:2001,Mandel:2003} where even individual imaging of single atoms has been accomplished \citep{Nelson2007, Bakr2009}. The current
challenge for all approaches is to scale the technology up for a larger number of qubits, for which several
proposals exist \citep{Kielpinski:2002,Duan2004,Cirac2000}.

The basic principle for quantum computation with trapped ions is
to use the internal electronic states of the ion as the qubit
carrier. Computational operations can then be performed by
manipulating the ions by coherent laser light
\citep{Haeffner:2008,Blatt:2008}. In order to perform entangling
gates between different ions, \citet{Cirac:1995} proposed to use
the mutual Coulomb interaction to realize a collective quantum bus
(a term used to denote an object that can transfer quantum
information between subsystems). The coupling between laser light and ion motion enables the
coherent mapping of quantum information between internal and
motional degrees of freedom of an ion chain. Two-ion gates are of particular
importance since combined with single-qubit rotations, they
constitute a universal set of quantum gates for computation
\citep{DiVincenzo:1995}. Several gate realizations have been
proposed
\cite{Cirac:1995,Molmer:1999,Milburn:2000,Jonathan:2000,Poyatos:1998,Minert:2001,Monroe:1997,Ripoll:2005}
and realized by several groups
\cite{Schmidt-Kaler:2003,Leibfried:2003,Kim:2008,Benhelm:2008,DeMarco:2002}.
When more ions are added to the ion chain, the same procedure can
be applied until the different vibrational-mode frequencies
become too close to be individually addressable\footnote{although there are
schemes that address multiple modes \citep{Kim:2009,Zhu:2006,Zhu:2006b}}; the
current state-of-the-art is the preparation and read-out of an
W entangled state of eight ions \cite{Haeffner:2005} and a
six-ion GHZ state \citep{Leibfried:2005}.

A way to solve this scalability problem is to use segmented
ion traps consisting of different regions between which the ions are shuttled (transported back and forth) \citep{Kielpinski:2002, Amini:2010}. The
proper optimization of the shuttling processes and optimization of
the laser ion interaction can only be fully performed with the aid
of numerical tools \citep{Huber:2010,Schulz:2006,Hucul:2008,Reichle:2006}. In
our presentation equal emphasis is put on the presentation of the physics of
quantum information experiments with ions and the basic ideas of
the numerical methods. All tools are demonstrated with ion trap
experiments, such that the reader can easily extend and apply the
methods to other fields of physics. Included is supplementary material, e.g.
source code and data such that even an inexperienced reader may
apply the numerical tools and adjust them for his needs. While
some readers might aim at understanding and learning the numerical methods by
looking at our specific ion trap example others might intend to
get a deeper understanding of the physics of quantum information
experiments through simulations and simulated
experiments. We start in Sec.~\ref{sec:IonTraps} with the
description of the ion trap principles and introduce numerical
methods to solve for the electrostatic potentials arising from the
trapping electrodes. Accurate potentials are needed to numerically
integrate the equation of motion of ions inside the trap.
Efficient stable solvers are presented in Sec.~\ref{sec:ClassicalTrajectories}. The axial motion of the ion is
controlled by changing the dc voltages of the electrodes. However,
usually we would like to perform the inverse, such that we find the
voltages needed to be applied to the electrodes in order to
produce a certain shape of the potential to place the ion at a
specific position with the desired trap frequency as described in
Sec.~\ref{sec:ClassicalControl}. This problem belongs to a type
of problems known as inverse problems, which are quite common in
physics. In Sec.~\ref{sec:TDSE} we enter the quantum world
where we first will obtain the stationary motional eigenstates of
the time-independent Schr\"odinger equation in arbitrary
potentials. We then describe methods to tackle the time-dependent
problem, and present efficient numerical methods to solve the time-dependent Schr\"odinger equation. The presented methods are used
in Sec.~\ref{sec:OCT} where we consider time-dependent
electrostatic potentials with the goal to perform quantum control
on the motional wavefunction and present the optimal control
algorithm. Finally, we apply these techniques in Sec.~\ref{sec:OCTgate} to the Cirac-Zoller gate. In the conclusion Sec.~\ref{sec:Conclusion}, we give a short account on the applicability of the presented numerical methods to qubit implementations other than trapped laser cooled ions.

\section{Ion trap development -- calculation of electrostatic fields}
\label{sec:IonTraps}
\begin{figure}
\includegraphics[width=0.50\textwidth,angle=0]{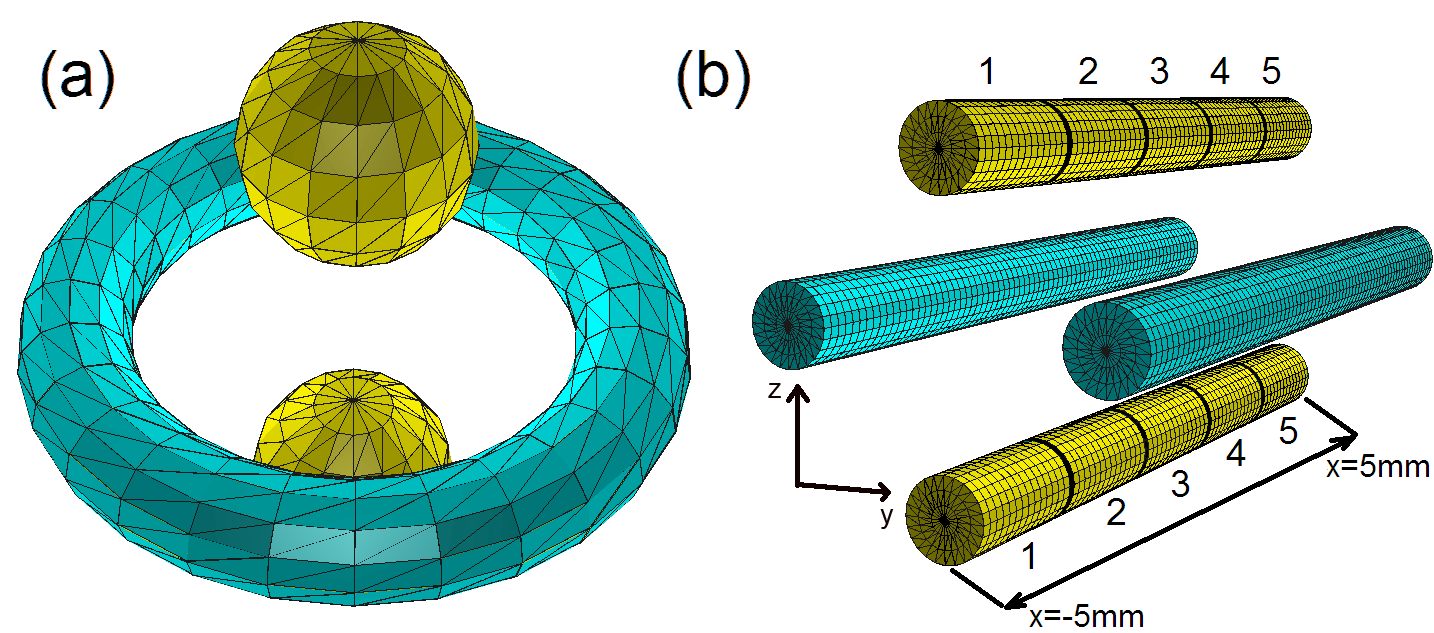}
\caption{(Color online). Electrode geometries of ion traps: The rf electrodes are
depicted in blue and dc electrodes in yellow respectively. (a)
Typical electrode configuration for a 3D ring trap with dynamic rf
confinement in all three dimensions. (b) Electrode arrangement for
a linear Paul trap. The dc electrodes are divided into segments
numbered from 1 to 5. For the numerical simulations we assume the following parameters: Segment have a width of 2~mm and a radius of 0.5~mm. The
central dc electrode is centered at the position $x=0$. The
minimum distance of the electrode surface to the trap axis is 1.5
mm. } \label{fig:ringtrap}
\end{figure}

\begin{figure}[tbp]
  \includegraphics[width=0.50\textwidth,angle=0]{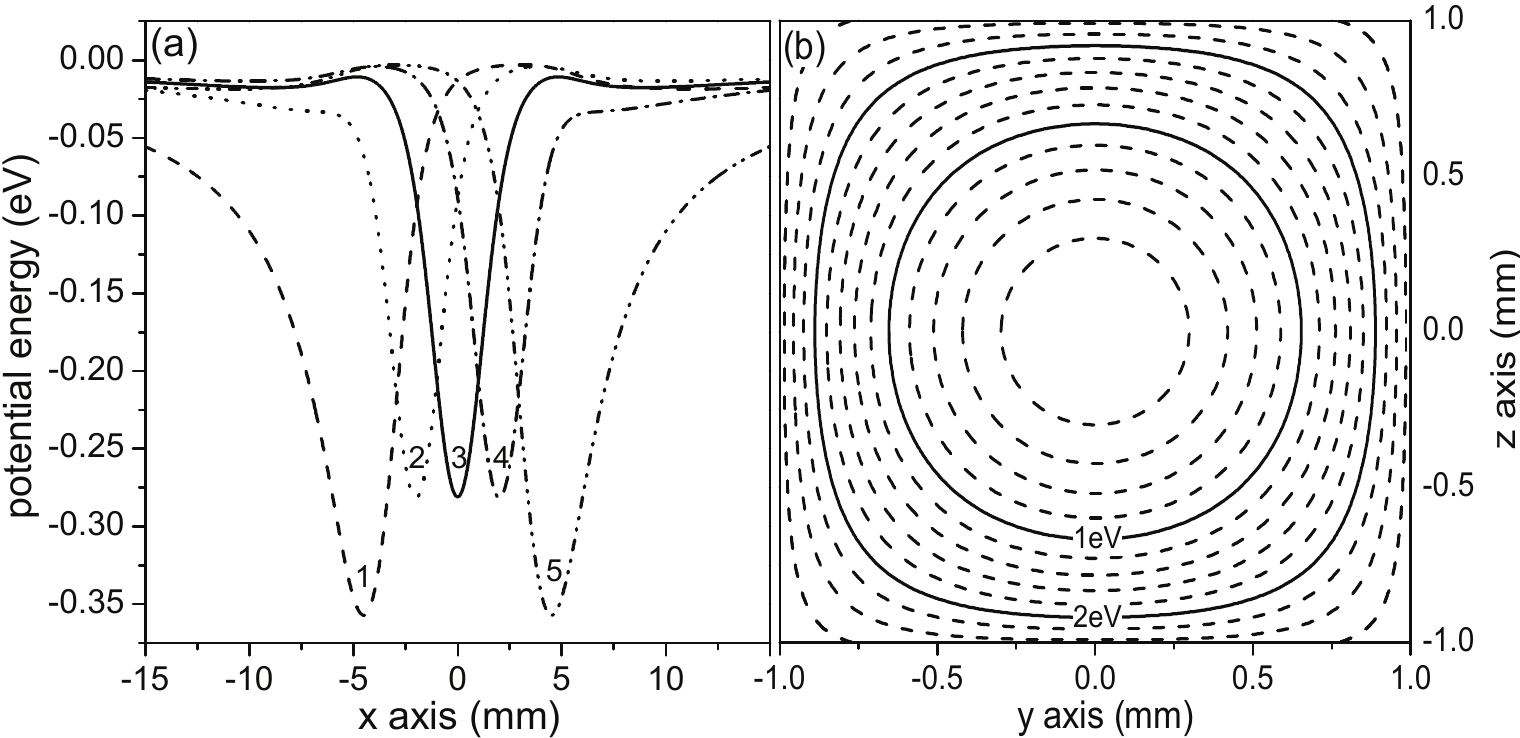}
\caption{ (a) Trapping potentials along the $x$-axis generated by
each individual electrode from the linear Paul trap geometry of
Fig.~\ref{fig:ringtrap}(b). Each curve corresponds to the respective
electrode biased to -1~V and all others to 0~V. (b) Equipotential
lines of the pseudo-potential in the radial plane
($U_\textrm{rf}=$200 V$_\textrm{pp}$,
$\omega_\textrm{rf}=2\pi\times 20$ MHz). Potentials are obtained
as described in Sec.~\ref{sec:IonTraps}. }
\label{fig:potentials}
\end{figure}

The scalability problem for quantum information with ion traps can be resolved with segmented ion traps. The trap potentials have to be tailored to control the position and trapping frequency of the ions. In the following, we will describe the mode of operation of a simple ion trap and then present numerical solvers for the efficient calculation of accurate electrostatic fields. Due to the impossibility of generating an electrostatic potential minimum in free space, ions may either be trapped in a combination of electric and magnetic static fields - a Penning trap \citep{Brown:1986}, or in a radio frequency electric field - a Paul trap, where a radio frequency (rf) voltage $U_\textrm{rf}$
with rf drive frequency $\omega_{\textrm{rf}}$ is applied to some of the
ion-trap electrodes \cite{Paul:1990}. In the latter case, we generate a potential
\begin{eqnarray}\nonumber
  \Phi(x,y,z,t)&=&\frac{U_{\textrm{dc}}}{2}(\alpha_\textrm{dc} x^2+ \beta_\textrm{dc} y^2+\gamma_\textrm{dc}
  z^2) \\
  &+&\frac{U_{\textrm{rf}}}{2}\cos(\omega_{\textrm{rf}} t)
  (\alpha_{\textrm{rf}} x^2+ \beta_{\textrm{rf}} y^2+\gamma_{\textrm{rf}}
  z^2),\label{potential}
\end{eqnarray}
where $U_\textrm{dc}$ is a constant trapping voltages applied to the
electrodes. The Laplace equation in free space $\Delta
\Phi(x,y,z)=0$ puts an additional constraint on the coefficients:
$\alpha_\textrm{dc}+\beta_\textrm{dc}+\gamma_\textrm{dc}=0$ and
$\alpha_{\textrm{rf}}+\beta_{\textrm{rf}}+\gamma_{\textrm{rf}}=0$.
One possibility to fulfill these conditions is to set
$\alpha_\textrm{dc}=\beta_\textrm{dc}=\gamma_\textrm{dc}=0$ and $
\alpha_{\textrm{rf}}+\beta_{\textrm{rf}}=-\gamma_{\textrm{rf}}$.
This produces a purely dynamic confinement of the ion and is
realized by an electrode configuration as shown in
Fig.~\ref{fig:ringtrap}(a), where the torus-shaped electrode is
supplied with radio frequency and the spherical electrodes are
grounded. An alternative solution would be the choice
$-\alpha_\textrm{dc}=\beta_\textrm{dc}+\gamma_\textrm{dc}$ and $
\alpha_{\textrm{rf}}=0,
\beta_{\textrm{rf}}=-\gamma_{\textrm{rf}}$, leading to a linear
Paul-trap with dc confinement along the $x$-axis and dynamic
confinement in the $yz$-plane. Fig.~\ref{fig:ringtrap}(b) shows a
possible setup with cylindrically shaped electrodes and segmented
dc electrodes along the axial direction which we will consider in
the following. In this trapping geometry, the ions can crystallize
into linear ion strings aligned along the $x$-axis. The classical
equation of motion for an ion with mass $m$ and charge $q$ is
$m\ddot{\vec{x}}=-q\nabla\Phi$, with $\vec{x}=(x,y,z)$ \cite{James:1998}. For a
potential given by Eq.~(\ref{potential}) the classical equations
of motion are transformed into a set of two uncoupled Mathieu
differential equations \citep{Leibfried:2003,Haeffner:2008}
\begin{equation}
  \frac{d^2u}{d\xi^2}+(a_{u}-2q_{u}\cos(2\xi
  ))u(\xi)=0 \quad u=y,z,\label{Mathieu}
\end{equation}
with $2\xi=\omega_{\textrm{rf}} t$. The Mathieu equation belongs to
the family of differential equations with periodic boundary
conditions and its solution is readily found in textbooks (for
example, \cite{Abramowitz:1964}). For a linear Paul-trap, the parameters
$a_{u}$ and $q_{u}$ in the $yz$-plane are given by
\begin{eqnarray}\nonumber
  q_{y}&=&\frac{2|q|U_{\textrm{rf}}\beta_{\textrm{rf}}}{m\omega_{\textrm{rf}}^{2}},\quad{a_{y}=-\frac{4|q|U_{\textrm{dc}}\beta_{\textrm{dc}}}{m\omega_{\textrm{rf}}^{2}}}, \\
  q_{z}&=&-\frac{2|q|U_{\textrm{rf}}\gamma_{\textrm{rf}}}{m\omega_{\textrm{rf}}^{2}},\quad{a_{z}=\frac{4|q|U_{\textrm{dc}}\gamma_{\textrm{dc}}}{m\omega_{\textrm{rf}}^{2}}}.
\end{eqnarray}
The solution is stable in the range $0\leq\beta_{u}\leq1$, where
$\beta_{u}=\sqrt{a_{u}+q_{u}^2/2}$ only depends on the parameters $a_{u}$ and $q_{u}$. The
solution of Eq.~(\ref{Mathieu}) in the lowest order approximation
($|a_{u}|, q_{u}^{2}\ll 1$), which implies that $\beta_{u}\ll 1$, is
\begin{equation}
  u(t)=u_{\textrm{0}}\cos(\omega_{u}t)\left(1+\frac{q_{u}}{2}\cos(\omega_{\textrm{rf}}t)\right).
\end{equation}
The ion undergoes harmonic oscillations at the
secular frequency $\omega_{u}=\beta_{u}\omega_{\textrm{rf}}/2$
modulated by small oscillations near the rf-drive frequency
(called micromotion). The static axial confinement along the
$x$-axis is harmonic with the oscillator frequency being given by
$\omega_{x}=\sqrt{|q|U_{\textrm{dc}}\alpha_{\textrm{dc}}/m}$. 
The axial confinement is generated by biasing the dc electrode
segments appropriately. Typical potential shapes can be
seen in Fig.~\ref{fig:potentials}(a).

The radial confinement is dominated by the rf
potential which can be approximated by an {\it effective}
harmonic potential $ \Phi_\textrm{eff}(y,z)=|q|
\left|\nabla \Phi(y,z) \right|^2/(4 m \omega_{\textrm{rf}}^2)$
where $\Phi(y,z)$ is the potential generated by setting the radio
frequency electrodes to a constant voltage $U_\textrm{rf}$ see
Fig.~\ref{fig:potentials}(b). However, this effective potential is
only an approximation and does not take the full dynamics of the
ion into account. Before we can simulate
the motion of the ion we need fast and accurate electrostatic
field solvers. In the next section we first present the
\emph{finite difference method} and then the \emph{finite element
method}. If the potentials need
to be known on a very small scale, a huge spatial grid would be
needed. Therefore, we introduce the \emph{boundary element method}
and show how the efficiency of this method can be drastically
improved by the application of the \emph{fast multipole method}.

\subsection{Finite difference method}
\label{sec:field} To obtain the electrostatic potential
$\Phi(x,y,z)$ in free space generated by a specific voltage
configuration $U_i$ for $i=1, \dots, n$ applied to the $n$
electrodes, we need to solve the Laplace equation $ \Delta \Phi
(x,y,z)=0$,
with the Dirichlet boundary condition $\Phi(x,y,z)=U_i$ for points
lying on the $i$th electrode. There are several approaches to
obtain the solution. The most intuitive is the finite difference
method (FDM). The principle is that we can write the differential
equation in terms of finite differences \cite{Thomas:1995}. To
illustrate this, we take the one dimensional differential equation
$\frac{d \Phi}{d x}=F(x)$ with the boundary condition
$\Phi(0)=a$ where $F(x)$ is an arbitrary function. If we
write
\begin{equation}
  \frac{d\Phi}{dx}=\lim_{\Delta x\rightarrow0} \frac{\Phi(x+\Delta x)-\Phi(x)}{\Delta x}=F(x),
\end{equation}
take only a finite difference $\Delta x$ and discretize the
$x$-axis by defining $x_i=i\,\Delta x$ with $i$ running from $0$
to $N$ and $x_N=1$, we obtain a discrete approximation which directly gives an explicit update equation (using the Euler
method) for $\Phi$:
\begin{equation}\label{eq:euler}
  \Phi(x_{i+1})=\Phi(x_i)+\Delta x F(x_i).
\end{equation}
Eq.~\eqref{eq:euler} can then be applied iteratively to solve the
differential equation. By comparing the solution with the Taylor
expansion and assuming that the higher order terms are bounded, we see that the error
of this finite difference approximation is of order $\Delta x$. This is
usually written as
\begin{equation}
  \frac{\Phi(x+\Delta x)-\Phi(x)}{\Delta x}= \frac{d\Phi}{dx} + \mathcal{O}(\Delta x),
\end{equation}
which means that there exists a constant $d$ such that
$\left|\frac{\Phi(x+\Delta x)-\Phi(x)}{\Delta x}-
\frac{d\Phi}{dx}\right|<d \left|\Delta x \right| $ for all $x$.
The Laplace equation is of second order, but one can
transform it into a set of a first order
differential equations
\begin{equation}
  \frac{d}{dx}
  \begin{pmatrix} \Phi \\ v \end{pmatrix} =
  \begin{pmatrix} v \\ F(x) \end{pmatrix},
\end{equation}
from which an explicit update rule can be derived, which is $\mathcal{O}(\Delta x)$. We can obtain a
second order approximation by cancelling the first order terms which gives a
centered-difference approximation for the first derivative
\begin{equation}
  \frac{\Phi(x_{n+1})-\Phi(x_{n-1})}{2 \Delta x}=\left.\frac{d\Phi}{dx}\right|_{x_n} + \mathcal{O}(\Delta
  x^2),
\end{equation}
which is of order $\mathcal{O}(\Delta x^2)$. A
centered-difference approximation for the second derivative reads
\begin{equation}
  \frac{\Phi(x_{n+1})-2\Phi(x_n)+\Phi(x_{n-1})}{\Delta x^2}=\left.\frac{d^2\Phi}{dx^2}\right|_{x_n} + \mathcal{O}(\Delta
  x^2),
\end{equation}
which is again of order $\mathcal{O}(\Delta x^2)$. The update rule
for the one-dimensional Laplace equation $\frac{d^2\Phi}{d x^2}=0$
thus has the form
\begin{equation}
  \Phi(x_{n+1})-2\Phi(x_n)+\Phi(x_{n-1})=0,
\end{equation}
which is an implicit expression, since the solution now has to be
obtained by solving a linear system of algebraic equations. We
have to specify two boundary conditions which we assume to be
$\Phi(x_0)=U_1$ and $\Phi(x_N)=U_2$, where $U_1$ and $U_2$ are the
voltages supplied at the boundaries. The matrix equation then has
the form
\begin{equation}
  \begin{pmatrix}
      -2 & 1 & 0 & \cdots & 0 \\
      1 & -2 & 1 & & \vdots \\
      0 &  1 &-2 & \ddots & 0\\
      \vdots & & \ddots &\ddots & 1 \\
      0 &  \cdots & 0 & 1 & -2
\end{pmatrix}
\begin{pmatrix}
  \Phi(x_1)\\
  \Phi(x_2)\\
  \Phi(x_3)\\
  \vdots\\
  \Phi(x_{N-1})
\end{pmatrix} = \begin{pmatrix}
    -U_1\\
    0\\
    0\\
    \vdots\\
    -U_2
  \end{pmatrix}.
\end{equation}
This equation has a tridiagonal form and can be most efficiently solved
by the Thomas algorithm \cite{Press:2007}\footnote{see package \textit{octtool}, function \textit{tridag}}. A sparse
matrix with more off-diagonal entries is obtained when the two- or
three-dimensional Laplace equation is treated in a similar fashion.  The
solution is then obtained either by simple Gaussian elimination or more
efficiently by iterative methods, such as the successive over relaxation
method (SOR)\cite{Press:2007} or the generalized minimum residual method (GMRES)
\cite{Saad:2003}.

One advantage of FDM is that it is easy to implement on a
uniform Cartesian grid. But in modeling three-dimensional geometries
one usually favors a triangular non-uniform mesh, where the
mesh spacing is spatially adapted to the local complexity of the
geometry structures; \textit{e.g.} it makes sense to use a finer mesh
near the edges.
\subsection{Finite element method}
\begin{figure}[tbp]
\includegraphics[width=0.4\textwidth,angle=0]{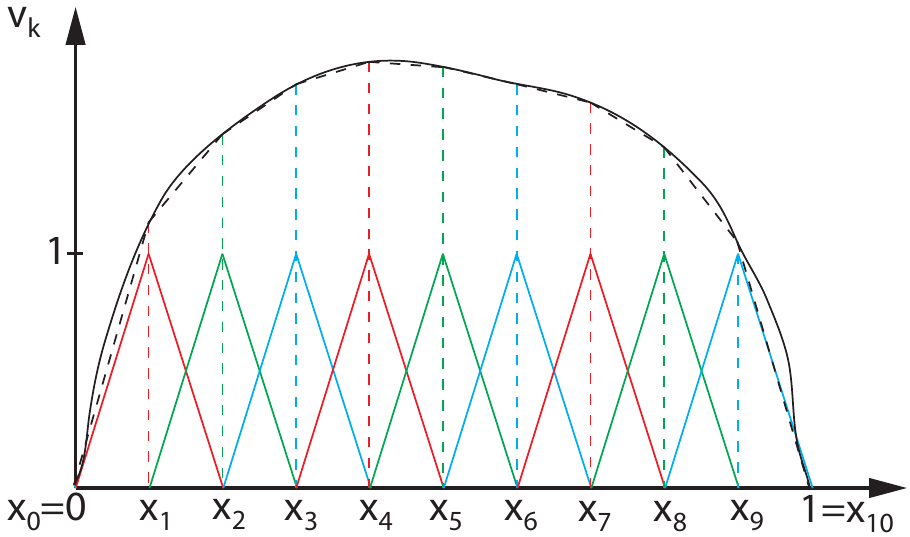}
\caption{(Color online). Overlapping basis functions from Eq.~(\ref{eq:basisv}) $v_k(x)$
(colored solid lines) for the finite element method providing linear interpolation (black dashed line) of an arbitrary function (black solid line).} \label{fig:vbasis}
\end{figure}

 The finite element method (FEM)
is better suited for nonuniform meshes with inhomogeneous granularity,
since it transforms the differential equation into an equivalent
variational one: instead of approximating the differential
equation by a finite difference, the FEM solution is
approximated by a finite linear combination of basis functions.
Again, we demonstrate the method with a one-dimensional
differential equation $\frac{d^2\Phi}{d x^2}=F(x)$, and for
simplicity we take the boundary condition $\Phi(0)=0$ and
$\Phi(1)=0$. The variational equivalent is an integral equation integrated between the boundaries at 0 and 1:
\begin{equation}
  \int_0^1 \frac{d^2\Phi(x)}{d x^2} v(x) d x \equiv \int_0^1
  F(x) v(x) d x,
\end{equation}
where $v(x)$ is the variational function which can be freely chosen except for the requirement $v(0)=v(1)=0$.
Integrating this by parts gives
\begin{eqnarray}
  \int_0^1 \frac{d^2\Phi(x)}{d x^2} v(x)
  d x&=&\underbrace{\left.\frac{d\Phi(x)}{d
        x}v(x)\right|_0^1}_{0}-\int_0^1 \frac{d\Phi(x)}{d
    x} \frac{d v(x)}{d x} d x \nonumber \\
  &\equiv& \int_0^1 F(x) v(x) d x. \label{eq:fem1}
\end{eqnarray}
We can now discretize this equation by constructing $v(x)$ on a
finite-dimensional basis. One possibility is linear interpolation:
\begin{equation}
  v_k(x)= \begin{cases}
    \frac{x-x_{k-1}}{x_k-x_{k-1}} & x_{k-1}\leq x\leq x_k, \\
    \frac{x_{k+1}-x}{x_{k+1}-x_k} & x_{k}< x\leq x_{k+1}, \\
    0 &  \text{otherwise},
\end{cases}\label{eq:basisv}
\end{equation}
with $x_0 = 0$, $x_N = 1$ and $x_k$ are the (not necessarily equidistant) sequential points in between and $k$ ranges from 1 to
$N-1$. These functions are shown in Fig.~\ref{fig:vbasis}. The advantage of this choice is that the inner products of
the basis functions $\int_0^1 v_k(x) v_j(x) d x$ and their
derivatives $\int_0^1 v'_k(x) v'_j(x) d x$ are only nonzero for
$|j-k|\leq 1$. The function $\Phi(x)$ and $F(x)$ are then
approximated by $\Phi(x)\approx \sum_{k=0}^N \Phi(x_k) v_k(x)$ and
$F(x)\approx \sum_{k=0}^N F(x_k) v_k(x)$ which linearly
interpolates the initial functions (see Fig.~\ref{fig:vbasis}).
With $\frac{d
  \Phi(x)}{d x} \approx \sum_{k=0}^N \Phi_k \frac{d
  v_k(x)}{d x}$, Eq.~\eqref{eq:fem1} is now recast into the form
\begin{multline}
-\sum_{k=0}^N \Phi(x_k) \left[ \int_0^1 \frac{d v_k(x)}{d
    x} \frac{d v_j(x)}{d x} \right] =\\ \sum_{k=0}^N
F(x_k) \left[\int_0^1 v_k(x) v_j(x) dx \right],
\end{multline}
where the terms in brackets are sparse matrices. This matrix equation
can then again be solved by iterative matrix solvers (such as GMRES).

For the Laplace problem, we need to extend this method to higher dimensions.  In
this case, instead of the integration by parts in Eq.~\eqref{eq:fem1} we
have to use Green's theorem \cite{Jackson:2009}:
\begin{align}
  \int_V \Delta \Phi(\vec{x}) v(\vec{x}) d V &=
  \underbrace{\int_{\delta V} \frac{\partial\Phi}{\partial n}v
    d s}_0-\int_V \nabla\Phi \nabla v d V \nonumber \\
  &\equiv \int_V F(\vec{x}) v(\vec{x}) d V, \label{eq:greenstheo}
\end{align}
where $V$ is the volume of interest and $\delta V$ the bounding
surface of the volume. Now space is discretized by three
dimensional basis functions and we can proceed in an analogous manner
as in the one dimensional case described above.

Potentials obtained by FDM and FEM usually result in unphysical
discontinuities (i.e.\ numerical artifacts) and must be smoothed in order
to be useful for ion trajectory simulations. Additionally, in order to obtain high accuracy trajectory simulations needed to simulate the trajectory extend of a trapped ion of less than 100 nm, the potentials that are calculated have to be interpolated, since computing with a grid with nanometer spacing would involve an unbearable computational overhead: the whole space
including the typically centimeter sized trap would have to be meshed
with a nanometer-spaced grid. FEM would allow for a finer mesh in the
region where the ion would be located reducing the overhead somewhat,
but this does not increase the accuracy of the surrounding coarser
grid. Avoiding to give a wrong expression we would like to stress that the FEM method finds wide applications in engineering and physics especially when complicated boundary conditions are imposed but for our accuracy goals FEM and FDM are inadequate.

\subsection{Boundary element method -- fast multipole method}
We proceed to show a different way of solving the Laplace problem
with a method which features a high accuracy and gives smooth
potentials that perform well in high-resolution ion-ray-tracing
simulations.

 To begin with, we divide the
electrodes into small surface elements $s_i$ of uniform surface charge
density $\sigma_i$, with $i$ numbering all surface elements
from $1$ to $N$. The potential at any point in space caused by a
charge distribution of these elements can be easily
obtained from Coulomb's law: one must simply sum up all the
contributions from each surface element. Hence the voltage $U_j$
on the surface element $s_j$ is generated by a linear
superposition of the surface charge densities $\sigma_i=\partial
\Phi(x_i)/\partial n$ also expressed as the normal derivative of the potential $\Phi$ (as obtained from the Maxwell equations) on all surface elements (including $s_j$)
additionally weighted by geometry factors given by the Coulomb law (represented by a matrix
$\hat{C}$) providing the following simple matrix equation
\begin{equation}
\label{eq:umq}
  \begin{pmatrix}
    U_1\\
    U_2\\
    \vdots\\
    U_N
  \end{pmatrix}
  =
  \hat{C}
  \begin{pmatrix}
    \sigma_1\\
    \sigma_2\\
    \vdots\\
    \sigma_N
  \end{pmatrix}.
\end{equation}
Now we want to solve for the surface charge densities in terms of the applied voltages.
The surface charge densities for a given voltage configuration can then be
obtained by finding the matrix inversion of $\hat{C}$. This is the
basic idea of the \textit{boundary element method} (BEM)
\cite{Pozrikidis:2002}. In the case of metallic surface elements where either the potential or the charge density is fixed, we have to exploit Green's second identity
\begin{equation}
  \Phi(\vec{x}_{j})=-2\sum_{i=1}^{N}\alpha_i(\vec{x}_{j})\frac{\partial \Phi(\vec{x}_{i})}{\partial n}+2\sum_{i=1}^N \beta_i(\vec{x}_{j})\Phi(\vec{x}_{i}) \label{eq:bem}.
\end{equation}
This equation has then to be solved for the unknown parameters which are the surface charge density $\frac{\partial \Phi(\vec{x}_{i})}{\partial n}$ on surfaces with given potential or the potential $\Phi(\vec{x}_{i})$ on surfaces with given charge density. Now we can choose the $\vec{x}_i$ and $\vec{x}_j$ to be representative points on each surface element \textit{e.g.} the center of gravity. This corresponds to the approximation that the potential and charge density are constant on each surface element. Eq.~(\ref{eq:bem}) is a matrix equation equivalent to Eq.~(\ref{eq:umq}). $\alpha_i$ is obtained by performing a surface integral
over the surface elements $s_i$ of the two-dimensional Green's
function $G(\vec{x},\vec{x}_{j})=-\frac{1}{2 \pi} \ln
\left|\vec{x}-\vec{x}_{j}\right|$ (for two-dimensional problems)
or the three-dimensional Green's function $G(\vec{x},\vec{x}_{j})=
\frac{-1}{4 \pi \left|\vec{x}-\vec{x}_{j}\right| }$ (for
three-dimensional problems). $\beta_i$ is obtained by performing a surface integral
over the surface elements $s_i$ of the gradient of the Green's
function multiplied by the surface norm $\vec{n}$
\begin{eqnarray}
  \alpha_i(\vec{x}_{j})=\oint_{s_i}G(\vec{x},\vec{x}_{j}) ds\label{eq:alpha},\\
  \beta_i(\vec{x}_{j})=\oint_{s_i}\vec{n}(\vec{x})\cdot \nabla G(\vec{x},\vec{x}_{j}) ds.\label{eq:beta}
\end{eqnarray}
Analytical expressions for these integrals over triangular surface
elements can be found in \citet{Davey:1989}, or via Gauss-Legendre
quadrature over a triangle. Eq.~\eqref{eq:bem} is now solved for the unknown parameters such as the surface charge densities
$\frac{\partial \Phi(\vec{x}_{i})}{\partial
  n}$. Once this is achieved, we can calculate the potential
\begin{equation}
  \Phi(\vec{x})=-\sum_{i=1}^{N}\alpha_i(\vec{x})\frac{\partial \Phi(\vec{x}_{i})}{\partial n}+\sum_{i=1}^N \beta_i(\vec{x})\Phi(\vec{x}_{i})\label{eq:fieldbem}
\end{equation}
at any position $\vec{x}$ with $\alpha_i(\vec{x})$ and
$\beta_i(\vec{x})$ evaluated at the same position. BEM is very
accurate and the implementation is quite straight-forward, but the
complexity of the matrix inversion scales prohibitively as
$\mathcal{O}(N^3)$. Different to the finite element method we cannot use sparse matrix solvers for this matrix inversion.

Fortunately, \citet{Greengard:1988} came up with an innovative method for
speeding-up the matrix vector multiplication needed for iterative matrix
inversion, which they termed the \emph{fast multipole method} (FMM). FMM
can solve the BEM problem with $\mathcal{O}(N)$ complexity, giving a
drastic increase in speed, and making BEM applicable to more complex
systems. In a series of publications, the algorithm was further improved
\citep{Carrier:1988,Greengard:1997,Cheng:1999,
  Nabors:1994,Gumerov:2005,Shen:2007} and extended to work with the
Helmholtz equation \cite{Gumerov:2004}. The basic idea was to use local
and far field multipole expansions together with efficient translation
operations to calculate approximations of the fields where the three-dimensional space is recursively subdivided into cubes. A detailed description of the method is beyond the scope of this paper and we refer to the cited literature.

\subsection{Application}
We have used the FMM implementation from \citet{Nabors:1994} and
combined it with a scripting language for geometry description and
the ability to read AutoCAD files for importing geometrical
structures. Any small inaccuracies due to numerical noise on the surface
charges are `blurred out' at large distances due to the Coulomb
law's $1/r$ scaling. In this regard, we can assert that the
surface charge densities obtained by FMM are accurate enough for our
purposes. If special symmetry properties are needed (such as
rotational symmetry for ion-lens systems or mirror symmetry) then
one can additionally symmetrize the surface charge densities. We have
implemented symmetrization functions in our code to support these
calculations \cite{Fickler:2009}. As FMM is used to speed up the matrix vector
multiplication it can be also used to speed up the evaluation of Eq.~(\ref{eq:fieldbem}) to obtain the potentials in free space. However, if accurate potentials in the sub micrometer scale are needed (such as for our application), it is better to use FMM for the
calculation of the surface charge densities \textit{i.e.} for the inversion of matrix Eq.~(\ref{eq:bem}) and then use conventional matrix
multiplication for the field evaluations as described by Eq.~(\ref{eq:fieldbem}).
Fig.~\ref{fig:potentials}(a) shows the smooth potentials
calculated by solving for the surface charge densities with FMM. Depicted
are the potentials for each electrode when biased to -1 V with all
others grounded. A trapping potential is then generated by taking
a linear superposition of these potentials.
Fig.~\ref{fig:potentials}(b) shows the equipotential lines of the
pseudo-potential. The full implementation can be found inside our
\textit{bemsolver} package together with example files for
different trap geometries.

With the calculated potentials from this chapter we can now solve
for the motion of an ion in the dynamic trapping potential of the
Paul trap which will be the focus of the next chapter.
\section{Ion trajectories -- classical equations of motion}
\label{sec:ClassicalTrajectories} The electrostatic potentials obtained with the methods presented in
the previous chapter are used in this section to simulate the trajectories of ions inside a dynamic trapping
potential of a linear Paul trap. We present the Euler method and
the more accurate Runge-Kutta integrators. Then we show that the
accuracy of trajectories can be greatly enhanced by using phase space
area conserving and energy conserving solvers such as the
\textit{St\"ormer-Verlet} method, which is a partitioned
Runge-Kutta integrator.
\subsection{Euler method}

The equation of motion of a charged particle with charge $q$ and mass
$m$ in an external electrical field can be obtained by solving the
ordinary differential equation
\begin{alignat}{2}\nonumber
  &&\ddot{\vec{x}}(t) & =\vec{f}(t,\vec{x}), \\ \nonumber
 \dot{\vec{y}}(t) \equiv&&
  \begin{pmatrix}
    \dot{\vec{x}}\\
    \dot{\vec{v}}
  \end{pmatrix}
  &=
\begin{pmatrix}
  \vec{v}\\
  \vec{f}(\vec{x},t)
\end{pmatrix}\equiv\vec{F}(t,\vec{y}),
\label{eq:motion}
\end{alignat}
where $\vec{f}(t,\vec{x})=(q/m) \vec{E}(t,x)=-(q/m) \nabla
\Phi(t,\vec{x})$ is the force arising from the electric field. The
vectors $\vec{y}$ and $\vec{F}$ are six-dimensional vectors
containing the phase space coordinates. As in the previous
section, the equation of motion can be solved by means of the
explicit Euler method with the update rule
$\vec{y}_{n+1}=\vec{y}_n+h \vec{F}(t_{n},\vec{y}_n)$, where we use
the notation $\vec{y}_{n}=\vec{y}(t_{n})$. If $\varepsilon$ is the
absolute tolerable error, then the time step $h=t_{n+1}-t_n$
should be chosen as $h=\sqrt{\varepsilon}$, which gives the best
compromise between numerical errors caused by the method and
floating point errors accumulated by all iterations.

An implicit variation of the Euler method
is given by the update rule $\vec{y}_{n+1}=\vec{y}_n+h
\vec{F}(t_{n+1},\vec{y}_{n+1})$. Neither of the methods is symmetric,
which means that under time inversion ($h \rightarrow -h$ and
$\vec{y}_n \rightarrow \vec{y}_{n+1}$), a slightly different
trajectory is generated. A symmetric update rule is given by the
implicit midpoint rule $\vec{y}_{n+1}=\vec{y}_n+h
\vec{F}((t_{n}+t_{n+1})/2,(\vec{y}_{n}+\vec{y}_{n+1})/2)$, which
has the additional property that it is \emph{symplectic}, meaning
that it is area preserving in phase space. The explicit and
implicit Euler methods are of order $\mathcal{O}(h)$ whereas the implicit
midpoint rule is of order $\mathcal{O}(h^2)$ \cite{Hairer:2002}.
These methods belong to the class of one stage \textit{Runge-Kutta} methods
\cite{Greenspan:2006}.

\subsection{Runge-Kutta method}
 The general $s$-stage Runge-Kutta method is
defined by the update equation
\begin{equation}
  \vec{y}_{n+1}=\vec{y}_n+h \sum_{i=1}^s b_i \vec{k}_i,
\end{equation}
with
\begin{align}
\label{eq:ki}
  \vec{k}_i&=\vec{F}(t_n+c_ih,\vec{y}_n+h\sum_{j=1}^s a_{ij}\vec{k}_j),\quad c_i =\sum_{j=1}^s a_{ij},
\end{align}
for $ i=1,\dots,s$ and $b_i$ and $a_{ij}$ are real numbers, which are given in \textit{Butcher tableaux} for several Runge-Kutta methods. Note that in the general case the $k_i$ are defined \textit{implicitly} such that Eq.~(\ref{eq:ki}) have to be solved at each time step. However, if
$a_{ij}=0$ for $i\leq j$ then the Runge-Kutta method is known as
\emph{explicit}. The standard
solver used in many numerical packages is the explicit 4th and 5th
order \textit{Dormand-Price} Runge-Kutta, whose values are given in
the Butcher tableau in Tab.~\ref{tab:butcher}.
\begin{table}

\begin{tabular}{c|ccccccc}
$0$ &$\leftarrow c_{1}$ \\[3pt]
$\frac{1}{5}     $ & $ \frac{1}{5}$ & \multicolumn{1}{l}{$\leftarrow a_{21}$}\\[3pt]
$\frac{3}{10}    $ & $ \frac{3}{40}$ & $  \frac{9}{40}$ & $\leftarrow a_{32}$\\[3pt]
$\frac{4}{5}     $ & $ \frac{44}{45}     $ & $   -\frac{56}{15}  $ & $ \frac{32}{9}$\\[3pt]
$\frac{8}{9}     $ & $ \frac{19372}{6561} $ & $ -\frac{25360}{2187} $ & $ \frac{64448}{6561} $ & $ -\frac{212}{729}$ & & $a_{76}$\\[3pt]
1                & $ \frac{9017}{3168} $ & $   -\frac{355}{33} $ & $   \frac{46732}{5247} $ & $ \frac{49}{176} $ & $ -\frac{5103}{18656}$ & $\downarrow$ & $b_7$\\[3pt]
1                & $ \frac{35}{384}    $ & $ 0                 $ & $ \frac{500}{1113}   $ & $   \frac{125}{192} $ & $ -\frac{2187}{6784} $ & $ \frac{11}{84}$ & $\downarrow$\\[3pt]
\hline\\[-8pt]
$b^{4th}$         & $ \frac{35}{384}    $ & $0$ & $\frac{500}{1113}$ & $\frac{125}{192}$ & $-\frac{2187}{6784}$ & $\frac{11}{84} $ & $0$\\[3pt]
$b^{5th}$ & $ \frac{5179}{57600}$ & $0$ & $\frac{7571}{16695}$ & $\frac{393}{640}$ & $-\frac{92097}{339200}$ & $\frac{187}{2100}$ & $\frac{1}{40}$
\end{tabular}
\caption{Butcher tableau for the 4th and 5th order Runge-Kutta methods: The left most column contains the $c_i$ coefficients, the last two rows under the separation line contain the $b_i$ coefficients to realize a 4th order or 5th order Dormand-Price Runge-Kutta. The $a_{ij}$ coefficients are given by the remaining numbers in the central region of the tableau. Empty entries correspond to $a_{ij}=0$.}
\label{tab:butcher}
\end{table}
The difference between the 4th and the 5th order terms can be used
as error estimate for dynamic step size adjustment.

\subsection{Partitioned Runge-Kutta method}
A significant improvement can be achieved by partitioning the dynamical variables into two groups \textit{e.g.} position and velocity coordinates, and to use two \textit{different} Runge-Kutta methods for their propagation. To illustrate this we are dealing with a classical non-relativistic particle of
mass $m$ which can be described by the following Hamilton function
$H(\vec{x},\vec{v})=T(\vec{v})+\Phi(\vec{x})$ with $T(\vec{v})=m
\vec{v}^2/2$. The finite
difference version of the equation of motion generated by this Hamiltonian reads $\vec{x}_{n+1}-2
\vec{x}_n +\vec{x}_{n-1}=h^2\vec{f}(\vec{x}_n)$. The only problem
is that we cannot start this iteration, as we do not know
$\vec{x}_{-1}$. The solution is to introduce the velocity
$\vec{v}=\dot{\vec{x}}$ written as a symmetric finite difference
\begin{eqnarray}
\vec{v}_{n}=\frac{\vec{x}_{n+1}-\vec{x}_{n-1}}{2h}
\end{eqnarray}
and the initial conditions: $\vec{x}(0)=\vec{x}_0,\,\dot{\vec{x}}(0)=\vec{v}_0$ such that we
can eliminate $\vec{x}_{-1}$ to obtain $\vec{x}_1=\vec{x}_0+h\vec{v}_0+\frac{h^2}{2}\vec{f}(\vec{x}_0)$. Now we can use the following recursion
relation:
\begin{eqnarray}
\vec{v}_{n+1/2}=\vec{v}_n+\frac{h}{2}\vec{f}(\vec{x}_n)\label{eq:v0.5},\\
\vec{x}_{n+1}=\vec{x}_n+h \vec{v}_{n+1/2},\\
\vec{v}_{n+1}=\vec{v}_{n+1/2}+\frac{h}{2} \vec{f}(\vec{x}_{n+1})\label{eq:v1}.
\end{eqnarray}
One should not be confused by the occurrence of half integer
intermediate time steps. Eqs. \eqref{eq:v0.5} and \eqref{eq:v1}
can be merged to $\textbf{v}_{n+1/2}=\textbf{v}_{n-1/2}+h
\textbf{f}(\textbf{x}_n)$ if $\textbf{v}_{n+1}$ is not of
interest. The described method is the \textit{St\"ormer-Verlet}
method and is very popular in molecular dynamics simulations where
the Hamiltonian has the required properties, \textit{e.g.} for a conservative system of $N$ particles with two-body interactions:
\begin{eqnarray}
H(\vec{x},\vec{v})=\frac{1}{2}\sum_{i=1}^N m_i \textbf{v}_i^T
\textbf{v}_i+\sum_{i=2}^N\sum_{j=1}^{i-1}V_{ij}(\left|\textbf{x}_i-\textbf{x}_j\right|).
\end{eqnarray}
In the case of the simulation of ion crystals, $V_{ij}$ would be the Coulomb interaction between ion pairs. The popularity of this method is due to the fact that it respects the conservation of energy and is a symmetric and symplectic solver of order two. It belongs to the general class of partitioned Runge-Kutta methods where the $s$-stage method for the partitioned differential equation $\dot{\vec{y}}(t)=\vec{f}(t,\vec{y},\vec{v})$ and $\vec{\dot{v}}(t)=\vec{g}(t,\vec{y},\vec{v})$ is defined by
\begin{table}
\begin{tabular*}{0.15\textwidth}{@{\extracolsep{\fill}}c|ccc}
$0$ & 0 & 0 &0\\[3pt]
$\frac{1}{2}     $ & $ \frac{5}{24}$ & $ \frac{1}{3}$&$ -\frac{1}{24}$\\[3pt]
$1   $ & $ \frac{1}{6}$ & $  \frac{2}{3}$ & $\frac{1}{6}$\\[3pt]
\hline\\[-8pt]
& $\frac{1}{6}$ & $\frac{2}{3}$ & $\frac{1}{6}$
\end{tabular*}\qquad
\begin{tabular*}{0.15\textwidth}{@{\extracolsep{\fill}}c|ccc}
$0$ & $\frac{1}{6}$ & $-\frac{1}{6}$ & $0$\\[3pt]
$\frac{1}{2}     $ & $ \frac{1}{6}$ & $ \frac{1}{3}$&$ 0$\\[3pt]
$1   $ & $ \frac{1}{6}$ & $  \frac{5}{6}$ & $0$\\[3pt]
\hline\\[-8pt]
& $\frac{1}{6}$ & $\frac{2}{3}$ & $\frac{1}{6}$
\end{tabular*}
\caption{Butcher tableau for the 4th order partitioned Runge-Kutta method consisting of a 3 stage Lobatto IIIA-IIIB pair \cite{Hairer:2002}: Left table shows the $b_i$,
$a_{ij}$ and $c_i$ coefficients corresponding to the table entries as described in the legend of Tab.~\ref{tab:butcher}. Right table shows corresponding primed $b_i'$,
$a_{ij}'$ and $c_i'$ coefficients.}
\label{tab:butcherPart}
\end{table}
\begin{eqnarray}
\vec{y}_{n+1}&=&\vec{y}_n+h \sum_{i=1}^s b_i \vec{k}_i,\\
\vec{v}_{n+1}&=&\vec{v}_n+h \sum_{i=1}^s b'_i \vec{l}_i,
\end{eqnarray}
\begin{eqnarray}
\vec{k}_i&=&\vec{f}(t_n+c_ih,\vec{y}_n+h\sum_{j=1}^s a_{ij}\vec{k}_j,\vec{v}_n+h\sum_{j=1}^s a'_{ij} \vec{l}_j),\nonumber\\
\vec{l}_i&=&\vec{g}(t_n+c'_ih,\vec{y}_n+h\sum_{j=1}^s a_{ij}\vec{k}_j,\vec{v}_n+h\sum_{j=1}^s a'_{ij} \vec{l}_j),\nonumber
\end{eqnarray}
with $b_i$ and $a_{ij}$ $(i,j=1,\ldots,s)$ being real numbers and
$c_i=\sum_{j=1}^s a_{ij}$ and analogous definitions for $b'_i$,
$a'_{ij}$ and $c'_i$. As an example Tab.~\ref{tab:butcherPart} shows the Butcher tableau for the 4th order partitioned Runge-Kutta method consisting of a 3 stage Lobatto IIIA-IIIB pair. A collection of more Butcher tableaux can be found in \cite{Hairer:2002} and \cite{Greenspan:2006}.
\begin{figure}
\includegraphics[width=0.5\textwidth,angle=0]{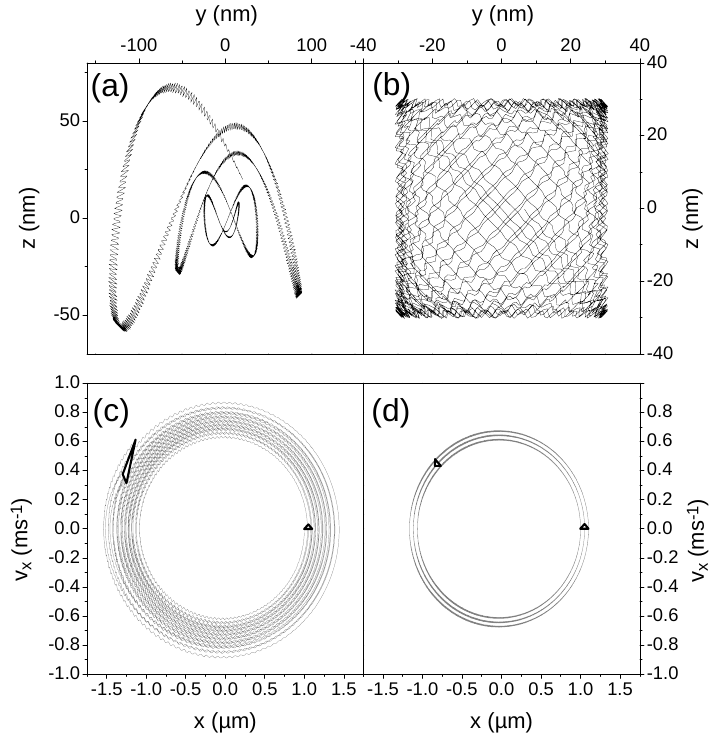}
\caption{Comparison of the Euler and St\"ormer-Verlet simulation methods: (a) Shows trajectories in the radial plane simulated with the
explicit Euler method. It can be clearly seen that the trajectories are unstable. (b) Simulation for the same parameters with the St\"ormer-Verlet method results in
stable trajectories. The small oscillations are due to the micro
motion caused by the rf drive. (c) Phase space trajectories of the harmonic
axial motion of an ion numerically integrated with the explicit
Euler method. The simulation was performed for a set of three different initial phase space coordinates, as indicated by the triangles. Energy and phase space area conservation are violated. (d) Equivalent trajectories
numerically integrated with the St\"ormer-Verlet method which
respects energy and phase space area conservation. The parameters are:
Simulation time 80 $\mu$s, number of simulation steps 4000,
$U_\textrm{rf}=$400 V$_\textrm{pp}$,
$\omega_\textrm{rf}=2\pi\times 12$ MHz,
$U_\textrm{dc}=(0,1,0,1,0)$V for the trap geometry shown in Fig.~\ref{fig:ringtrap}(b). } \label{fig:trajectory}
\end{figure}
\subsection{Application}
We have simulated the phase space trajectories of ions in the
linear Paul trap of Fig.~\ref{fig:ringtrap}(b) for 500,000 steps
with the Euler method and with the St\"ormer-Verlet method. A trajectory in the radial plane obtained with the Euler method can be seen in Fig.~\ref{fig:trajectory}(a), it clearly shows an unphysical instability. When simulated by the St\"ormer-Verlet method, the
trajectories are stable, as can be seen in
Fig.~\ref{fig:trajectory}(b). The small oscillations are due to
the micro motion caused by the rf drive. The Euler method does not lead to results obeying to energy and  phase space area conservation (see Fig.~\ref{fig:trajectory}(c)), whereas the
St\"ormer-Verlet method conserves both quantities (see
Fig.~\ref{fig:trajectory}(d)). The Euler method should be avoided
when more than a few simulation steps are performed. The
St\"ormer-Verlet integrator is implemented in the
\textit{bemsolver} package.

In the next section we will find out how we can control the
position and motion of the ion.

\section{Transport operations with ions -- ill-conditioned inverse problems }
\label{sec:ClassicalControl}
\begin{figure}
\includegraphics[width=0.40\textwidth,angle=0]{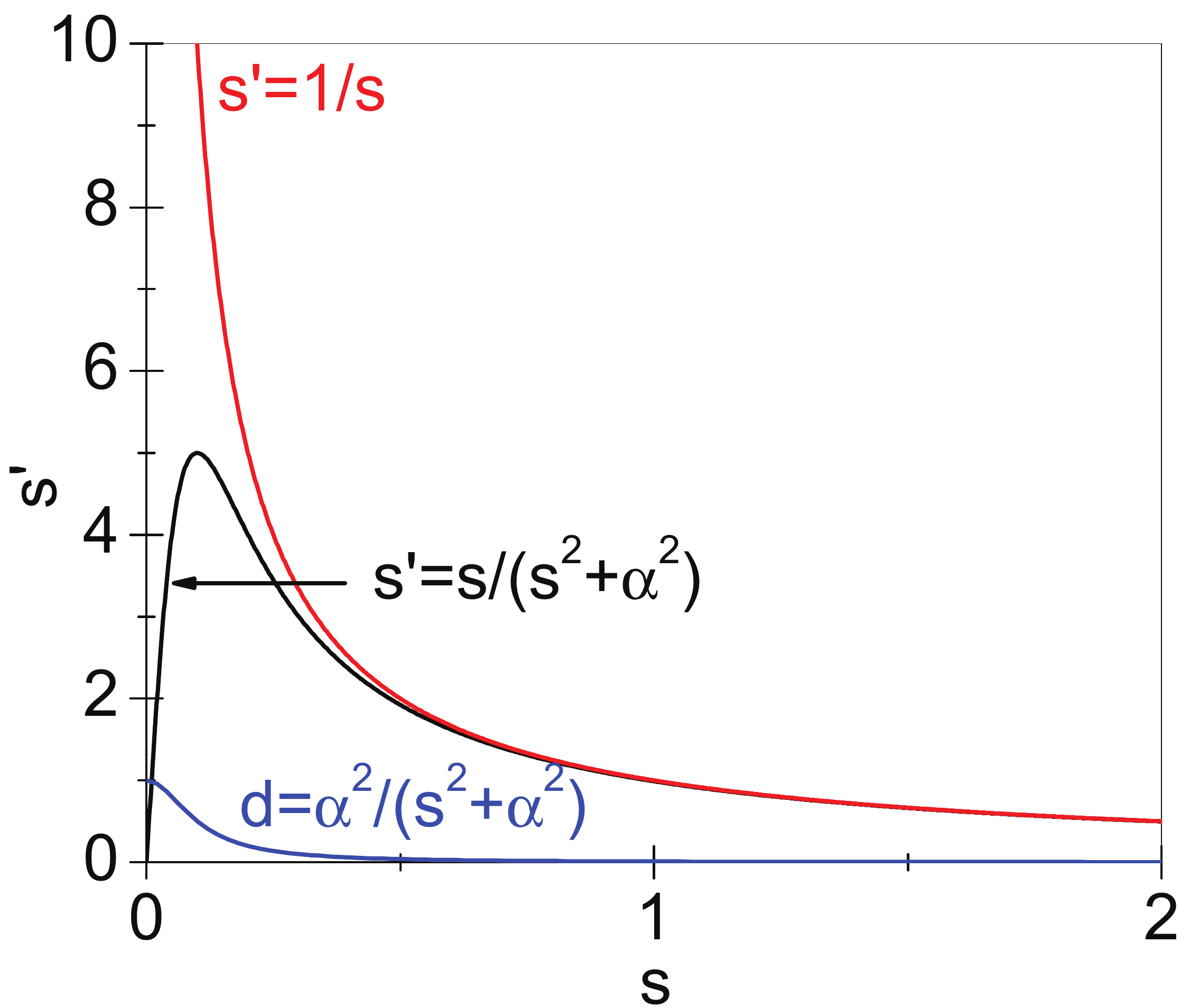}
\caption{(Color online). Illustration of the regularization technique: Suppression of the divergence at zero (red curve) of the $1/s$ term. The black curve shows the Tikhonov regularization term, the singular behavior of the $1/s$ inverse is avoided and diverging values are replaced by values near zero. The blue curve tends towards one for $s \rightarrow 0$ where $1/s$ is diverging. It is used to force diverging inverses towards a given value (see text).} \label{fig:tikhonov}
\end{figure}
If we wish to transport ions in multi-electrode geometries, the electrostatic potential has
to be controlled. An important experimental constraint is that the
applicable voltage range is always limited. Additionally, for the
dynamic change of the potentials, voltage sources are generally
band limited and therefore we need the voltages of the electrodes
to change smoothly throughout the transport process. The starting point for the solution of this problem is to consider $A(x_i,j)=A_{ij}$, a unitless matrix representing the potential on all points $x_i$ when each electrode $j$ is biased to 1V whereas all other electrodes are kept at 0V (see Fig.~\ref{fig:potentials}(a)). Hence we can calculate the generated total
potential $\Phi(x_i)=\Phi_i$ at any position $x_i$ by the linear
superposition
\begin{equation}
\Phi_i=\sum_{j=1}^{N}
A_{ij}U_j,\quad{i=1,\ldots,M},
\label{eq:PhiAU}
\end{equation}
with $N$ denoting the number of separately controllable electrodes
and $M$ being the number of grid points in space, which could be
chosen, for example, on the trap axis. We would like to position the
ion at a specific location in a harmonic potential with a desired
curvature, \textit{i.e.} trap frequency. This is a matrix inversion problem, since we have
specified $\Phi_i$ over the region of interest but we need to find the voltages
$U_j$. The problem here is that $M$ is much larger than $N$, such that the matrix $A_{ij}$ is over determined, and (due to some unrealizable
features or numerical
artifacts) the desired potential $\Phi(x_i)$ might not lie in the
solution space. Hence a usual matrix inversion, in most cases, will
give divergent results due to singularities in the inverse of
$A_{ij}$.

This class of problems is called \textit{inverse problems}, and
if singularities occur then they are called \textit{ill-conditioned}
inverse problems.

In our case, we wish to determine the electrode voltages for a
potential-well moving along the trap axis, therefore a series of
inverse problems have to be solved. As an additional constraint we require that
the electrode-voltage update at each transport step is limited. In the
following, we will describe how the \textit{Tikhonov
regularization} method can be employed for finding a matrix inversion avoiding singularities and fulfilling additional constraints\cite{Press:2007,Tikhonov:1977}.
\subsection{Thikonov regularization}
For notational simplicity, we will always assume that $A$ is a $M \times N$ dimensional matrix, the potentials along the trap axis are given by the vector
\begin{equation}
\vec{\Phi}=(\Phi(x_{1}),\Phi(x_{2}),\ldots,\Phi(x_{M}))^{T}
\end{equation}
and $\vec{u}=(U_{1},U_{2},\ldots,U_{N})^{T}$ is a vector containing the electrode voltages. Instead of
solving the matrix equation $A \vec{u} =\vec{\Phi}$, the Thikonov method minimizes the residual
 $\left|\left|A \vec{u}-\vec{\Phi}\right|\right|^2$
with respect to the Euclidean norm. This alone could still lead to diverging values for some components of $\vec{u}$ which is cured by imposing an
additional minimization constraint through the addition of a regularization term such as
\begin{equation}
\alpha\left|\left|\vec{u}\right|\right|^2,
\label{eq:alpha1}
\end{equation}
 which penalizes diverging values. The larger the real valued weighting parameter $\alpha$ is chosen, the more this penalty is weighted and large values in $\vec{u}$ are suppressed at the expense that the residual might increase. Instead of resorting to numerical iterative minimizers a faster and more deterministic method is to perform a \textit{singular-value decomposition} which decomposes the $M \times N$
matrix $A$ into a product of three matrices $A=U S V^T$,
 where $U$ and $V$
are unitary matrices of dimension $M\times M$ and $N \times N$, and $S$ is a diagonal (however not quadratic) matrix with diagonal
entries $s_i$ and dimension $M\times N$. Singular-value decomposition routines are contained
in many numerical libraries, for example
\textsc{lapack}\footnote{\texttt{http://www.netlib.org/lapack}}. The
inverse is then given by the $A^{-1}=V S'U^T$ where
$S'=S^{-1}$. The diagonal entries of $S'$ are related to those of $S$ by
$s'_i=1/s_i$ such that the singular terms can now be
directly identified. The advantage of the singular value
decomposition now becomes clear: all the effects of singularities are contained only in the diagonal matrix $S'$. By addressing these singularities (i.e., when $s_i \rightarrow 0$) in the proper way, we avoid any divergent behavior in the voltages $U_j$. The easiest way to deal with the
singularities would be to set all terms $s'_i$ above a certain threshold
to zero. Thikonov however uses the smooth truncation function
\begin{equation}
s'_i=\frac{s_i}{s_i^2+\alpha^2}
\label{eq:si}
\end{equation}
 (see Fig.~\ref{fig:tikhonov} black curve)
which behaves like $1/s_i$ for large $s_i$ but tends to zero for
vanishing $s_i$, providing a
gradual cutoff. The truncation parameter $\alpha$ has the same meaning as above in the regularization term: the larger $\alpha$ the more the diverging values are forced towards zero, and if $\alpha=0$ then the exact inverse will be calculated and diverging values are not suppressed at all.
The required voltages are now obtained by
\begin{equation}
\label{eq:inverse}
\vec{u}=V S' U^T\vec{\Phi}.
\end{equation}
These voltages fulfill the requirement to lie within some given technologically accessible voltage range, which can be attained by iteratively adjusting $\alpha$.

In the remainder of this section, we present an extension of this method which is better adapted to our specific task of smoothly shuttling an ion between two trapping sites: instead of generally minimizing the electrode voltages, we would rather like to limit the \textit{changes in the voltage} with respect to the voltage configuration $\vec{u_0}$ which is already applied prior to a single shuttling step. Therefore the penalty function of Eq.~(\ref{eq:alpha1}) is to be replaced by
\begin{equation}
\alpha\left|\left|\vec{u}-\vec{u_0}\right|\right|^2.
\end{equation}
The application of this penalty is achieved through an additional term in
Eq.~(\ref{eq:inverse})
\begin{equation}
\vec{u}=V S' U^T \vec{\Phi}+ V D V^T\
\vec{u_0}.
\label{eq:thikonew}
\end{equation}
$D$ is a $N \times N$ diagonal matrix with entries
$d_i=\frac{\alpha^2}{s_i^2+\alpha^2}$ which shows an opposite behavior as the Thikhonov truncation function of Eq.~(\ref{eq:si}): where the truncation function leads to vanishing voltages avoiding divergencies, $d_i$ tends towards one (see Fig.~\ref{fig:tikhonov} blue curve) such that the second term in Eq.~(\ref{eq:thikonew}) keeps the voltages close to $\vec{u_0}$. By contrast for large values of $s_i$ the corresponding values of $d_i$ are vanishing as $s_i^{-2}$ such that the second term in Eq.~(\ref{eq:si}) has no effect. The $N \times N$ matrix $V$ is needed to transform the vector $\vec{u_0}$ into the basis of the matrix $D$.

A remaining problem now arises when \textit{e.g.} a single electrode, \textit{i.e.} a column in Eq.~(\ref{eq:PhiAU}) leads to a singularity. This might due to the fact that it does generate only small or vanishing fields at a trap site of interest. Even though the singularity is suppressed by the regularization it nevertheless leads to a global reduction of all voltages which adversely affects the accuracy of the method. This can be resolved by the additional introduction of weighting factors $0<w_j<1$ for each electrode such that the matrix $A$ is replaced by $A'_{ij}=A_{ij} w_j$. The voltages $u'_j$ obtained from Eq.~(\ref{eq:thikonew}) with accordingly changed matrices $U$, $V$, $S'$ and $D$ are then to be rescaled $1/w_j$. A reasonable procedure would now be to start with all $w_j=1$ and to iteratively decrease the $w_j$ for electrodes for which the voltage $u_j$ is out of range.

\subsection{Application}
\begin{figure}
\includegraphics[width=0.50\textwidth,angle=0]{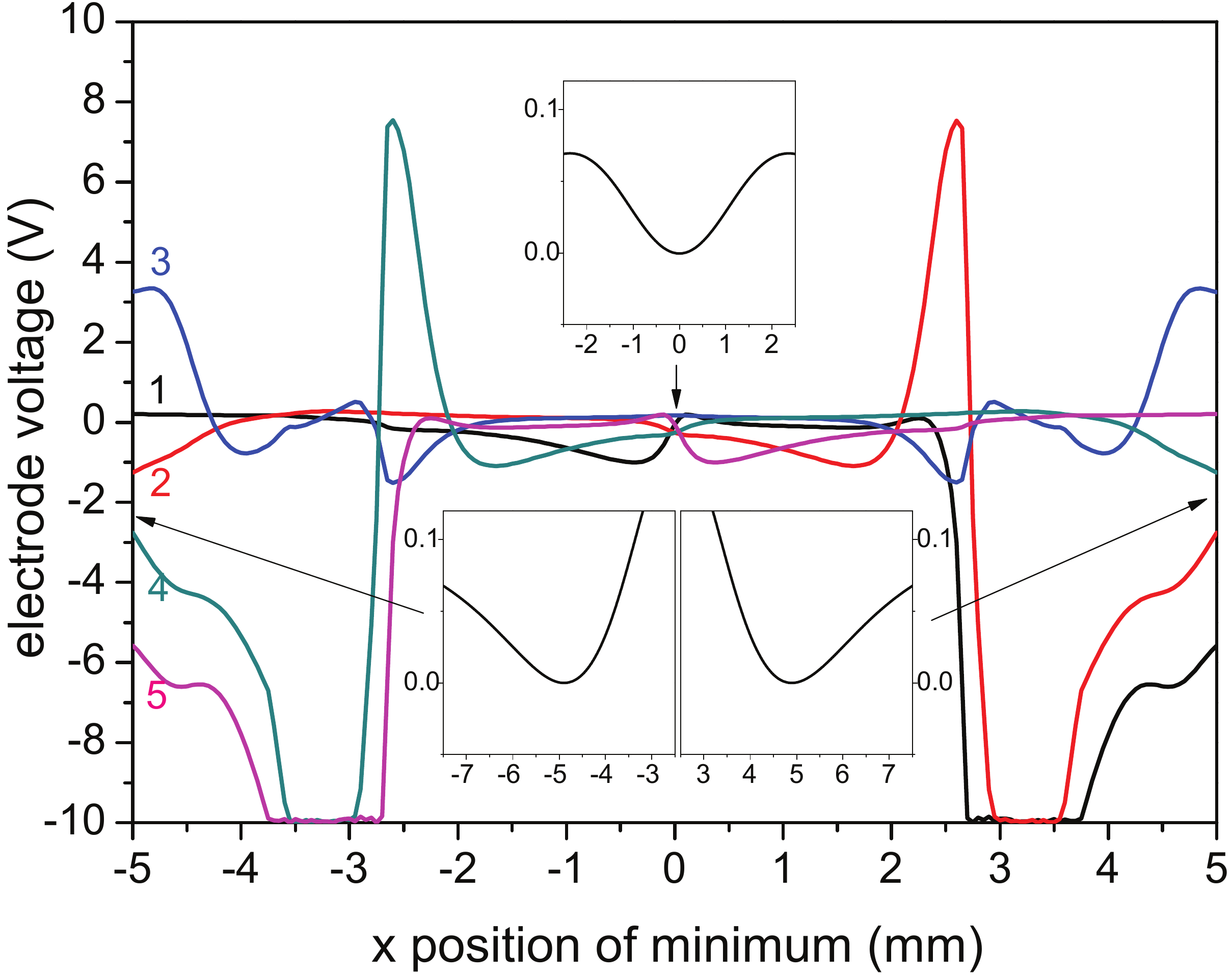}
\caption{(Color online). Voltage configurations to put the minimum of a harmonic potential with fixed curvature into a different positions. The insets show the resulting potentials obtained by linear superposition of the individual electrode potentials.} \label{fig:regularized}
\end{figure}

 A full implementation of the algorithm can be found in the supplied numerical tool box in the \textit{svdreg} package. Fig.~\ref{fig:regularized} shows the obtained voltages which realize a harmonic trapping potential $\Phi(x)=\delta (x-x_0)^2$ with $\delta=0.03 V/\textrm{mm}^2$ at different positions $x_0$ with a given voltage range of $-10\leq U_i \leq 10$ for the linear five-segmented trap of Fig.~\ref{fig:ringtrap}(b). We have also experimentally verified the accuracy of the potentials by using the ion as a local field probe with sub-percent agreement to the numerical results \cite{Huber:2010}.

\section{Quantum dynamics -- efficient numerical solution of the Schr\"odinger equation }
\label{sec:TDSE} As the trapped atoms can be cooled close to the
ground state, their motional degrees of freedom have to be
described quantum mechanically. In this chapter we present methods
to solve the time-independent Schr\"odinger equation and the time-dependent Schr\"odinger equation. The presented tools are used in
the Sec.~\ref{sec:OCT} and \ref{sec:OCTgate} about optimal
control. For trapped ions, the relevance of treating the quantum dynamics of the motional degrees is not directly obvious as the trapping potentials are extremely harmonic, such that (semi)classical considerations are often fully sufficient. However, full quantum dynamical simulations are important for experiments outside the Lamb-Dicke regime \cite{McDonnel:2007,Poschinger:2010}, and for understanding the sources of gate infidelities \cite{Kirchmair:2009}. For trapped neutral atoms, the confining potentials are generally very anharmonic such that quantum dynamical simulations are of fundamental importance.

\subsection{Solution of the time-independent Schr\"odinger equation -- the Numerov method}
\label{sec:TISE}
The stationary eigenstates for a given external potential are solutions of the time-independent Schr\"odinger equation (TISE)
\begin{equation}
\left(
-\frac{\hbar^{2}}{2m}\frac{d^2}{dx^2}-\Phi(x)\right)\psi(x)=E
\psi(x). \label{eq:tise}
\end{equation}
For the harmonic potentials generated with the method of the
previous chapter these solutions are the harmonic oscillator
eigenfunctions. But how can we obtain the eigenfunctions and
eigenenergies for an arbitrary potential? A typical textbook
solution would be choosing a suitable set of basis functions and
then diagonalizing the Hamiltonian with the help of linear algebra packages such as \textsc{lapack} to obtain the eigenenergies and
the eigenfunctions as linear combination of the basis functions.

\begin{figure}
\includegraphics[width=0.50\textwidth,angle=0]{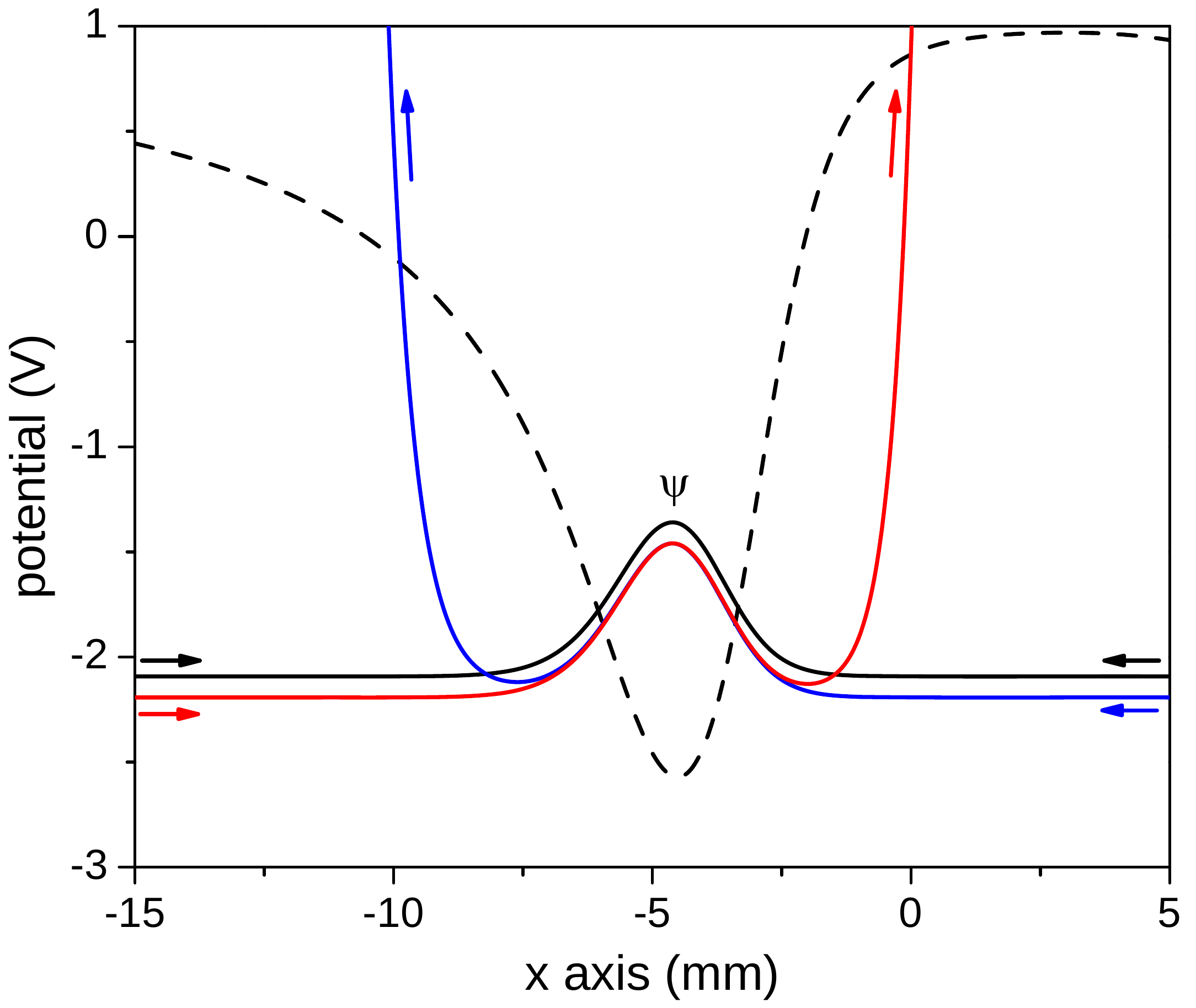}
\caption{(Color online). Illustration of the Numerov algorithm for the numerical solution of the TISE: The dashed line shows the trapping potential for a particle. The black solid line shows the ground state eigenfunction. Blue and red lines show the result of numerical integration starting from right and left respectively if the energy does not correspond to an energy eigenvalue.} \label{fig:numerov}
\end{figure}

A simple approach is exploiting the fact that physical
solutions for the wavefunction have to be normalizable. This condition for the wavefunction leads to the
constraint that the wavefunction should be zero for $x \rightarrow
\pm \infty$. Thus it can be guessed from the potential shape where
the nonzero parts of the wavefunction are located in space, and
Eq.~\eqref{eq:tise} can be integrated from a starting point
outside this region with the eigenenergy as an initial guess
parameter. For determining the correct energy eigenvalues, we make
use of the freedom to start the integration from the left
or from the right of this region of interest.  Only if correct
eigenenergies are chosen as initial guess, the two wavefunctions
will be found to match (see Fig.~\ref{fig:numerov}). This
condition can then be exploited by a root-finding routine to
determine the proper eigenenergies. If the Schr\"odinger equation
is rewritten as $\frac{d^2}{dx^2}\psi(x)=g(x)\psi(x)$, then the
Numerov method \cite{Blatt:1967} can be used for integration.
Dividing the $x$-axis into discrete steps of length $\Delta x$,
the wavefunction can be constructed using the recurrence relation
\begin{eqnarray}
\psi_{n+1}&=&\frac{\psi_n(2+\frac{10}{12}g_n\Delta x^2)-\psi_{n-1}(1-\frac{1}{12}g_{n-1}\Delta x^2)}{1-\frac{\Delta x^2}{12}g_{n+1}}\nonumber\\&&+{\mathcal O}(\Delta x^6),
\end{eqnarray}
where $\psi_n = \psi(x + n \Delta x)$, $g_n = g(x + n \Delta x)$.
With this method, the stationary energy eigenstates are obtained.
The source code is contained in the \textit{octtool} package.

\subsection{Numerical evaluation of the time-dependent Hamiltonian}
In order to understand the behavior of quantum systems under the influence of external
control field and to devise strategies for their control, we perform numerical
simulations of the time evolution. In the case of systems with very few degrees of freedom,
the task is simply to solve linear first order differential equations
and/or partial differential equations, depending on the representation
of the problem. Already for mesoscopic systems, the Hilbert space
becomes so vast that one has to find suitable truncations to its regions
which are of actual relevance. Here, we will only deal with the simple case of
only one motional degree of freedom and possible additional internal degrees
of freedom.

An essential prerequisite for the propagation of quantum
systems in time is to evaluate the Hamiltonian; first, one must find its appropriate matrix representation, and second, one
needs to find an efficient way to describe its action on a given
wavefunction. The first step is decisive for finding the
eigenvalues and eigenvectors, which is often important for a
meaningful analysis of the propagation result, and the second step
will be necessary for the propagation itself. We assume that we
are dealing with a particle without any internal degrees of freedom moving along one spatial dimension $x$. A further assumption is
that the particle is to be confined to a limited portion of
configuration space $0\leq x \leq L$ during the time interval
of interest. We can then set up a homogeneous grid
\begin{equation}
  x_i = i\;\Delta x, \ i=1,\dots,N,\ \Delta x = \frac{L}{N}.
  \label{eq:gridsetup}
\end{equation}
A suitable numerical representation of the wavefunction is given by a set of $N$ complex
numbers
\begin{equation}
  \psi_i=\psi(t,x_i).
  \label{eq:discretepsirepresentation}
\end{equation}
The potential energy part of the Hamiltonian is diagonal in position
space, and with $V_i=V(x_i)$ it is straightforwardly applied to the
wavefunction:
\begin{equation}
  \hat{V}\psi(x) \rightarrow V_i \psi_i.
  \label{eq:potentialenergyonwfn}
\end{equation}
One might now wonder how many grid points are necessary for a
faithful representation of the wavefunction. The answer is given
by the \textit{Nyquist-Shannon sampling theorem}. This theorem
states that a band limited waveform can be faithfully reconstructed
from a discrete sampled representation if the sampling period is
not less than half the period pertaining to the highest frequency
occurring in the signal. Returning to our language, we can
represent the wavefunction \textit{exactly} if its energy is
limited and we use at least one grid point per antinode. Of
course, one still has to be careful and consider the possible
minimum distance of antinodes for setting up a correct grid.
Eq.~\eqref{eq:discretepsirepresentation} then gives an exact
representation, and Eq.~\eqref{eq:potentialenergyonwfn} becomes an
equivalence.

The kinetic energy operator, however, is \emph{not} diagonal in position
space, because the kinetic energy is given by the variation of the
wavefunction along the spatial coordinate, i.e.\ its second derivative:
\begin{equation}
  \hat{T}=-\frac{\hbar^2}{2m} \frac{d^2}{d x^2}.
\end{equation}
One could then apply $\hat{T}$ by means of finite differences
(see Sec.~\ref{sec:IonTraps}) which turns out to be extremely
inefficient as one would have to use very small grid steps (large
$N$) in order to suppress errors. At the very least, we would have to be
sure that the grid spacing is much smaller than the minimum
oscillation period, which is in complete contrast to the sampling
theorem above. In order to circumvent this problem, we consider
that $\hat{T}$ is diagonal in momentum representation, with
the matrix elements
\begin{equation}
\tilde{T}_{nn'}=\langle k_n|\hat{T}| k_{n'} \rangle=\frac{\hbar^2
    k_n^2}{2m}\delta_{nn'}.
\end{equation}
Thus, we can directly apply the kinetic energy operator on the
wavefunction in momentum space:
\begin{equation}
  \hat{T}\tilde{\psi}(k) \rightarrow \tilde{T}_{ii} \tilde{\psi}_i ,\quad i=1,\dots,M,
\end{equation}
where $\tilde{\psi}(k)$ is the momentum representation of the wavefunction.

The quantity we need is the position representation of $\psi(x)$
with the kinetic energy operator applied to it, which gives
\begin{eqnarray}
  \left(\hat{T}\psi\right)_l&=&\langle x_l|\hat{T}|\psi\rangle \nonumber \\
  &=&\sum_{j=1}^{N} \langle x_l |\hat{T}| x_j \rangle \langle x_j|\psi\rangle \nonumber \\
  &=&\sum_{j=1}^{N}\sum_{n=1}^{M} \langle x_l |k_n\rangle \langle k_n |\hat{T}|k_n\rangle \langle k_n| x_j \rangle \langle x_j|\psi\rangle \nonumber \\
  &=&\frac{1}{M}\sum_{n=1}^{M} e^{ik_n x_l} \frac{\hbar^2 k_n^2}{2m} \sum_{j=1}^{N} e^{-i k_n x_j} \psi_j \nonumber \\
  &=&\sum_{j=1}^{N}T_{lj}\psi_{j},
  \label{eq:crucialkineticenergyequation}
\end{eqnarray}
where $\mathcal{F}_{nj}=\langle k_{n}| x_{j}\rangle
=e^{-ik_{n}x_{j}}/\sqrt{M}$. An explicit expression for the matrix
$T_{lj}$ will be given below. After addition of the diagonal
potential energy matrix one obtains the total Hamiltonian in the
position representation, it then can be diagonalized by means of
computer algebra programs or efficient algorithms as the
\texttt{dsyevd} routine of the computational algebra package
\textsc{lapack}.

An interesting perspective on the propagation problem is seen when
the second last line of
Eq.~\eqref{eq:crucialkineticenergyequation} is read from right to
left, which gives a direct recipe for the efficient application of
the kinetic energy operator:

\begin{enumerate}
\item Transform the initial wavefunction to momentum space by performing
  the matrix multiplication
  \begin{equation}
    \tilde{\psi}_n = \sum_{j=1}^{N} \mathcal{F}_{nj} \psi_j.
    \label{eq:propagationstep1}
  \end{equation}

\item Multiply with the kinetic energy matrix elements:
  \begin{equation}
    \tilde{\psi}'_n =\tilde{T}_{nn} \tilde{\psi}_n.
    \label{eq:propagationstep2}
  \end{equation}

\item Transform back to position space by performing another matrix
  multiplication
  \begin{equation}
    (\hat{T}\psi)_l=\sum_{n=1}^{M} \mathcal{F}^{\ast}_{ln}
    \tilde{\psi}'_n.
    \label{eq:propagationstep3}
  \end{equation}

\end{enumerate}
These three steps can, of course, be merged into only one matrix
multiplication, but the crucial point here is to notice that the matrix
multiplications are nothing more than a \emph{Discrete Fourier
  Transform} (DFT), which can be performed on computers with the
\emph{Fast Fourier Transform} algorithm (FFT) \cite{Cooley1965}. This has
the tremendous advantage that instead of the $N^2$ scaling for matrix
multiplication, the scaling is reduced to $N\log N$.

Up to here, we have made no statement on how the grid in momentum space
defined by the $k_n$ and $M$ is to be set up. The usage of FFT
algorithms strictly requires $M=N$. The grid in momentum space is then
set up by
\begin{equation}
  k_n = n\,\Delta k , \ -\frac{N}{2}+1\leq n \leq \frac{N}{2} ,\ \Delta k = \frac{2K}{N},
  \label{eq:kgridsetup}
\end{equation}
analogously to Eq.~\eqref{eq:gridsetup}. The maximum kinetic
energy is then simply $T_{\text{max}}=\tfrac{\hbar^2K^2}{2m}$. The
remaining free parameter for a fixed position space grid
determined by $L$ and $N$ is then the maximum wavenumber $K$. Its
choice is motivated by the sampling theorem: the position space
step $\Delta x$ is to be smaller than the minimum nodal distance
$\lambda_{\text{min}}/2=\pi/K$, which establishes the relation
\begin{equation}
\Delta x=\beta\frac{\pi}{K}=\beta\frac{2\pi}{\Delta kN}.
\label{eq:deltaxdeltarelation}
\end{equation}
$\beta$ is a `safety factor' which is to be chosen slightly
smaller than one to guarantee the fulfillment of the sampling
theorem. For the sake of clarity, we note that Eq.~(\ref{eq:deltaxdeltarelation}) is equivalent to
\begin{equation}
\frac{L}{N}=\beta\frac{\pi}{K}. \label{eq:KLrelation}
\end{equation}
The optimum number of grid points $N_{\text{opt}}$ for efficient
but accurate calculations is then determined by energy
conservation, i.e. the grid should provide a maximum possible
kinetic energy $T_{\text{max}}$ equal to the maximum possible
potential energy $V_{\text{max}}$. The latter is determined by the
specific potential pertaining to the physical problem under
consideration, whereas $T_{\text{max}}$ is directly given by the grid
step. We can therefore state:
\begin{eqnarray}
V_{\text{max}}&=&T_{\text{max}}=\frac{\hbar^2 K^2}{2m} \nonumber \\
&=&\frac{1}{2m}\left(\frac{\beta\pi\hbar N}{L}\right)^2 \nonumber \\
\Rightarrow N_{\text{opt}} &= & \frac{L}{\beta\pi\hbar} \sqrt{2m
V_{\text{max}}}.
\end{eqnarray}
If more grid points are chosen, the computational effort increases
without any benefit for the accuracy. In turn the results become
inaccurate for fewer grid points. If we consider a harmonic
oscillator with $V(x)=\tfrac{1}{2}m\omega^2
\left(x-\tfrac{L}{2}\right)^2$, we obtain
$V_{\text{max}}=\tfrac{1}{8}m\omega^2 L^2$ and therefore
\begin{equation}
N_{\text{opt}}=\frac{m\omega L^2}{\beta h},
\end{equation}
which is (for $\beta=1$) exactly the number of eigenstates sustained
by the grid.

Now, the recipe for the application of the kinetic energy operator can be
safely performed, while in Eqs.~\eqref{eq:propagationstep1} and
\eqref{eq:propagationstep3} the matrix multiplications are simply to be
replaced by forward and backward FFTs respectively\footnote{Two caveats for
handling FFTs shall be mentioned. First, some FFT algorithms do
not carry out the normalization explicitly, such that one has to
multiply the resulting wavefunction with the correct normalization factor
after the backward FFT. Second, one is initially often confused by the
way the data returned by the FFT is stored: the first component typically
pertains to $k=0$, with $k$ increasing by $+\Delta k$ with increasing
array index. The negative $k$ components are stored from end to
beginning, with $k=-\Delta k$ as the last array element, with $k$
changing by $-\Delta k$ with decreasing array index.}. This \emph{Fast
  Fourier Grid Method} was first presented by \citet{Feit1982} and
\citet{Kosloff1983}. For a review on propagation schemes, see
\citet{Kosloff1988}.

With the grid having been set up, we are in good shape to
calculate the eigenstates of a given Hamiltonian by matrix
diagonalization as mentioned above. The explicit expression for
the position space matrix elements of Eq.~(\ref{eq:crucialkineticenergyequation}) is \citep{Tannor2007}
\begin{equation}
  T_{lj}=\frac{\hbar^2}{2m}
  \begin{cases}
    \frac{K^{2}}{3} \left(1+\frac{2}{N^2}\right) & \text{for } l=j\text{,}\\
    \frac{2K^2}{N^2} \frac{(-1)^{j-l}}{\sin^2(\pi (j-l)/N)}& \text{otherwise }  \text{.}\\
  \end{cases}
\end{equation}

The efficiency of the Fourier method can be significantly
increased for problems with anharmonic potentials, as is the case
with the Coulomb problem in molecular systems. If one is looking at
the classical trajectories in phase space, these will have
distorted shapes, such that only a small fraction of phase space
is actually occupied by the system. This leads to an inefficient
usage of the grid, unless an inhomogeneous grid is used. The
Nyquist theorem can be invoked \emph{locally}, such that the local
de Broglie wavelength $\lambda_{dB}=2\pi\left[2m
\left(E_0-V(x)\right)\right]^{-1/2}$ is to be used as the grid
step. Further information can be found in Refs. \cite{Fattal1996,
Kokoouline1999, Willner2004}.

The general propagator for time-dependent Hamiltonians is given by
\begin{equation}
  \hat{U}(t,t_0)=\hat{\mathcal{T}} e^{-\frac{i}{\hbar}\int_{t_0}^t \hat{H}(t')\;dt'},
\end{equation}
with the time ordering operator $\hat{\mathcal{T}}$ and a
Hamiltonian consisting of a static kinetic energy part and a
time-dependent potential energy part, which one may write as
\begin{equation}
  \hat{H}(t)=\hat{T}+\hat{V}(t).
\end{equation}
We
discretize the problem by considering a set of intermediate times
$t_n$, which are assumed to be equally spaced. The propagator is
then given by
\begin{equation}
  \hat{U}(t,t_0)= \hat{\mathcal{T}}  \prod_n e^{-\frac{i}{\hbar} \hat{H}(t_n)\;\Delta t}.
\label{eq:timediscretepropagator}
\end{equation}
It is important to state that the time-ordering operator now just keeps
the factors ordered, with decreasing $t_n$ from left to right.  Here,
the first approximation has been made by replacing the control field by
its piecewise constant simplification $\hat{V}(t_n)$. The short term
propagators in the product of Eq.~\eqref{eq:timediscretepropagator} have
to be subsequently applied to the initial wavefunction. The remaining
problem with the application of the short time propagators then arises due
to the non-commutativity of $\hat{T}$ and $\hat{V}(t_n)$. A possible way
out would be the diagonalization of $\hat{T}+\hat{V}(t_n)$ in matrix
representation, which is highly inefficient due to the unfavorable
scaling behavior of matrix diagonalization algorithms. Two main
solutions for this problem are widely used, namely the split-operator
technique and polynomial expansion methods, which are to be explained in
the following.

\subsection{The split-operator method}
The basic idea of the \textit{split-operator method} is to simplify the
operator exponential by using the product
\begin{equation}
  e^{-\frac{i}{\hbar} \hat{H}(t_n)\;\Delta t} \approx e^{-\frac{i}{2\hbar} \hat{V}(t_n)\;\Delta t} e^{-\frac{i}{\hbar} \hat{T}\;\Delta t} e^{-\frac{i}{2\hbar}\hat{V}(t_n)\;\Delta t}
\end{equation}
at the expense of accuracy due to violation of the non-commutativity of the
kinetic and potential energy operators \cite{Tannor2007}. The error scales as
$\Delta t^3$ if $\hat{V}(t_n)$ is taken to be the averaged potential over the time interval $\Delta t$ \cite{Kormann:2008} with $t_n$ being the midpoint of the interval. Additional complexity arises if one is
dealing with internal degrees of freedom, such that distinct states are
coupled by the external control field. This is exactly the case for light-atom or
light-molecule interaction processes. In these cases no diagonal representation of
$\hat{V}$ exists in position space, meaning that it has to be
diagonalized.

\subsection{The Chebyshev propagator}
A very convenient way to circumvent the problems associated with the
split-operator method is to make use of a polynomial expansion of the
propagator
\begin{equation}
  \exp\left( -\frac{i}{\hbar}\int_{t_0}^t \hat{H}(t^\prime)\ud t'\right) = \sum_{k} a_k\;\mathcal{H}_k(\hat{H}),
\label{eq:polynomialpropagatorexpansion}
\end{equation}
and exploit the properties of these polynomials. As we
will see, an expansion in terms of Chebyshev polynomials
$\mathcal{H}_k$ leads to a very favorable convergence behavior and
a simple implementation due to the recurrence relation
\begin{equation}
\label{eq:recrel}
  \mathcal{H}_{k+1}(x)=2x\,\mathcal{H}_{k}(x)-\mathcal{H}_{k-1}(x),
\end{equation}
with $\mathcal{H}_0(x)=1$ and $\mathcal{H}_1(x)=x$.
As Chebyshev polynomials are defined on an interval $x
\in\left[-1,1\right]$, the energy has to be mapped on this interval by shifting and rescaling:
\begin{equation}
  \hat{H}'=2\frac{\hat{H}-E_{<} \mathrm{1\!l}}{E_{>}-E_{<}}- \mathrm{1\!l},
\end{equation}
where $\mathrm{1\!l}$ is the unity matrix, and $E_{>}$ and $E_{<}$ denote the maximum and minimum eigenvalues of
the unscaled Hamiltonian, respectively. The propagation scheme is then
as follows:

\begin{enumerate}

\item Given the initial wavefunction $\psi(t=t_i)$, set
  \begin{eqnarray}
    \phi_0&=&\psi(t=t_i), \nonumber \\
    \phi_1&=&-i\,\hat{H}'\phi_0.
  \end{eqnarray}

\item Calculate
  \begin{equation}
    \phi_{n+1}=-2i\,\hat{H}'\phi_n+\phi_{n-1},
  \end{equation}
  for all $n<n_{\mathrm{max}}$, which is the recursion relation Eq.~(\ref{eq:recrel}) applied on the wavefunction.

\item Sum the final wavefunction according to
  \begin{equation}
    \psi(t_n+\Delta t)=e^{-\frac{i}{2 \hbar}(E_{<}+E_{>})\Delta t}\sum_{n=0}^{n_{\mathrm{max}}} a_n \phi_n.
  \end{equation}

\end{enumerate}

The phase factor in front of the sum corrects for the energy rescaling and the expansion coefficients are given by Bessel functions:
\begin{equation}
  a_n=
  \begin{cases}
    J_0\left(\frac{(E_{>}-E_{<})\Delta t}{2\hbar}\right) & \text{for } n=0\text{,}\\
    2 (-i)^n\;J_n\left(\frac{(E_{>}-E_{<})\Delta t}{2\hbar}\right) & \text{for } n>0 \text{.}\\
  \end{cases}
\end{equation}
It is interesting to note that the Chebyshev polynomials are not used
explicitly in the scheme. Due to the fact that the Bessel functions $J_n(z)$
converge to zero exponentially fast for arguments $z>n$, the accuracy of the propagation is only limited by
the machine precision as long as enough expansion coefficients are used. For time-dependent Hamiltonians however the accuracy is limited by the finite propagation time steps $\Delta t$, which should be much smaller than the time scale at which the control fields are changing. A detailed account on the accuracy of the Chebyshev method for time-dependent problems is given in \cite{Peskin:1994,Ndong:2010}. The suitable number of expansion coefficients can
easily be found by simply plotting their magnitudes. The most common
error source in the usage of the propagation scheme is an incorrect
normalization of the Hamiltonian. One has to take into account that the eigenvalues of the
Hamiltonian might change in
the presence of a time-dependent control field. A good test if the scheme is working at all is
to initialize the wavefunction in an eigenstate of the static
Hamiltonian and then check if the norm is conserved and the system stays in the initial state upon propagation
with the control field switched off. Another important point is that the
propagation effort is relatively independent of the time step $\Delta t$. For larger time steps, the number of required expansion coefficients
increases linearly, while the total number of steps decreases only. For
extremely small steps, the computational overhead of the propagation
steps will lead to a noticeable slowdown. All presented numerical propagators are contained in the \textit{octtool} package.

\section{Optimizing wavepacket manipulations -- optimal control theory (OCT)}
\label{sec:OCT} Now that we know how to efficiently simulate wavepackets in our quantum system and how to manipulate the potentials, we can begin to think about \emph{designing} our potentials to produce a desired evolution of our wavepacket, whether for transport, or for more complex operations (such as quantum gates). In this chapter, we discuss one method for achieving this in detail, namely optimal control theory \citep{Peirce1988, Kosloff1989, Tannor1992,Somloi1993, Krotov1996, Zhu1998, Sklarz2002, Khaneja2005, Krotov2008}. These methods belong to a class of control known as open-loop, which means that we specify everything about our experiment beforehand in our simulation, and then apply the results of optimal control directly in our experiment. This has the advantage that we should not need to acquire constant feedback from the experiment as it is running (an often destructive operation in quantum mechanics). We will focus on one particular method prevalent in the literature, known as the Krotov algorithm \citep{Tannor1992, Krotov1996, Somloi1993}.

\subsection{Krotov algorithm}
Optimal Control Theory (OCT) came about as an extension of the classical calculus of variations subject to a differential equation constraint. Techniques for solving
such problems were already known in the engineering community for some
years, but using OCT to optimize quantum mechanical systems only began
in the late 1980s with the work of Rabitz and coworkers \citep{Shi1988,
  Peirce1988}, where they applied these techniques to numerically obtain
optimal pulses for driving a quantum system towards a given goal. At
this time, the numerical approach for solving the resulting set of
coupled differential equations relied on a simple gradient method with
line search. In the years that followed, the field was greatly expanded
by the addition of more sophisticated techniques that promised improved
optimization performance. One of the most prominent amongst these is the
Krotov method \citep{Tannor1992, Krotov1996, Somloi1993} developed by
Tannor and coworkers for problems in quantum chemistry around the
beginning of the 1990s, based on Krotov's initial work. This method
enjoyed much success, being further modified by Rabitz \citep{Zhu1998} in
the late 90s.

Until this point, OCT had been applied mainly to problems in quantum
chemistry, which typically involved driving a quantum state to a
particular goal state (known as state-to-state control), or maximizing
the expectation value of an operator. The advent of quantum information
theory at the beginning of the new millennium presented new challenges
for control theory, in particular the need to perform quantum gates,
which are not state-to-state transfers, but rather full unitary
operations that map whole sets of quantum states into the desired final
states \citep{Palao2002, Calarco:2004, Hsieh2008}. \citet{Palao2002}
extended the Krotov algorithm to deal with such unitary evolutions,
showing how the method can be generalized to deal with arbitrary numbers
of states. This will become useful for us later when we want to optimize
the Cirac-Zoller gate. Other methods for optimal control besides Krotov
have also been extensively studied in the literature, most notably
perhaps being GRAPE \cite{Khaneja2005}, but these methods will not be
discussed here.
\paragraph{Constructing the optimization objective}
We will now proceed to outline the basics of optimal control theory as
it is used for the optimization later in this paper \footnote{For a tutorial on quantum optimal control
theory which covers the topics presented here in more detail, see
\citet{Werschnik2007}.}. We always begin by
defining the \emph{objective} which is a mathematical description of the
final outcome we want to achieve. For simplicity, we shall take as our
example a state-to-state transfer of a quantum state $\ket{\psi(t)}$
over the interval $t \in [0,T]$. We begin with the initial state
$\ket{\psi(0)}$, and the evolution of this state takes place in
accordance with the Schr\"{o}dinger equation
\begin{equation}
  \label{eq:schroedinger}
  i \hbar \frac{\partial}{\partial t} \ket{\psi(t)} = \hat{H}(t) \ket{\psi(t)},
\end{equation}
where $\hat{H}$ is the Hamiltonian. (Note that we will often omit
explicit variable dependence for brevity.) Now assume that the
Hamiltonian can be written as
\begin{equation}
  \label{eq:ham_sum}
  \hat{H}(t) = \hat{H}_0(t) + \sum_i \varepsilon_i(t) \hat{H}_i(t),
\end{equation}
where $H_0$ is the \emph{uncontrollable} part of the Hamiltonian
(meaning physically the part we cannot alter in the lab), and the
remaining $H_i$ are the \emph{controllable} parts, in that we may affect
their influence through the (real) functions $\varepsilon_i(t)$, which
we refer to interchangeably as `controls' or `pulses' (the latter
originating from the early days of chemical control where interaction
with the system was performed with laser pulses). Let's take as our goal
that we should steer the initial state into a particular final state at
our final time $T$, which we call the goal state $\ket{\psig}$. A
measure of how well we have achieved the final state is given by the
\emph{fidelity}
\begin{equation}
  \label{eq:fidelity}
  J_1[\psi] \equiv - | \bracket{\psig}{\psi(T)} |^2,
\end{equation}
which can be seen simply as the square of the inner product between the
goal state and the final evolved state. Note that $J_1[\psi]$ is a
\emph{functional} of $\psi$.

The only other constraint to consider in our problem is the dynamical
one provided by the time-dependent Schr\"{o}dinger equation (TDSE) in
Eq.~\eqref{eq:schroedinger}. We require that the quantum state must
satisfy this equation at all times, otherwise the result is clearly
non-physical. If a quantum state $\ket{\psi(t)}$ satisfies
Eq.~\eqref{eq:schroedinger}, then we must have
\begin{equation}
  \left(\partial_t + \tfrac{i}{\hbar}\hat{H}\right)\ket{\psi(t)} = 0,\ \forall t \in T, \quad
  \text{where} \quad
  \partial_t = \frac{\partial}{\partial t}.
\end{equation}
We can introduce a Lagrange multiplier to cast our constrained
optimization into an unconstrained one. Here, we introduce the state
$\ket{\chi(t)}$ to play the role of our Lagrange multiplier, and hence
we write our constraint for the TDSE as
\begin{align}
  \nonumber
    J_2[\varepsilon_i, \psi, \chi] &\equiv \int_0^T
    \left(\bra{\chi(t)}(\partial_t +
      \tfrac{i}{\hbar}\hat{H})\ket{\psi(t)} + \mathrm{c.c.}\right)\ud t
  \\ \label{eq:tdse_const}
  & = 2 \mathrm{Re} \int_0^T \bra{\chi(t)}(\partial_t +
      \tfrac{i}{\hbar}\hat{H})\ket{\psi(t)}\ud t,
\end{align}
where we have imposed that both $\ket{\psi(t)}$ and
$\bra{\psi(t)}$ must satisfy the TDSE.

\paragraph{Minimizing the objective}
Now that we have defined our goal and the constraints, we can write our
\emph{objective} $J[\varepsilon_i, \psi, \chi]$ as
\begin{equation}
  \label{eq:objective}
  J[\varepsilon_i, \psi, \chi] = J_1[\psi] + J_2[\varepsilon_i, \psi, \chi].
\end{equation}
The goal for the optimization is to find the minimum of this functional
with respect to the parameters $\psi(t)$, $\chi(t)$ and the controls
$\varepsilon_i(t)$. In order to find the minimum, we consider the
stationary points of the functional $J$ by setting the total variation
$\delta J = 0$. The total variation is simply given by the sum of the
variations $\delta J_\psi$ (variation with respect to $\psi$), $\delta
J_\chi$ (variation with respect to $\chi$), and $\delta
J_{\varepsilon_i}$ (variation with respect to $\varepsilon_i$), which we
set individually to zero. For our purposes, we define the variation of a
functional
\begin{equation}\label{eq:fd_def}
  \delta_\psi F[\psi] = F[\psi + \delta \psi] - F[\psi].
\end{equation}
This can be thought of as the change brought about in $F$ by perturbing
the function $\psi$ by a small amount $\delta \psi$.

Considering $\delta J_\psi$, we have
\begin{equation}
  \label{eq:dj_psi} \nonumber
  \delta J_\psi = \delta J_{1,\psi} + \delta J_{2,\psi}.
\end{equation}
Using our definition from Eq.~\eqref{eq:fd_def} for $\delta J_{1,\psi}$
results in
\begin{align}\nonumber
  \delta J_{1,\psi} &= J_1[\psi + \delta \psi] - J_1[\psi] \\ \nonumber
  &= -|\bracket{\psig}{\psi(T) + \delta \psi(T)}|^2 +
  |\bracket{\psig}{\psi(T)}|^2 \\
  \begin{split}
    \nonumber &= -\bracket{\psi(T) + \delta \psi(T)}{\psig}\bracket{\psig}{\psi(T) + \delta \psi(T)} \\
    & \qquad + |\bracket{\psig}{\psi(T)}|^2
  \end{split}\\
  \begin{split}
    \nonumber &= - \bracket{\delta
      \psi(T)}{\psig}\bracket{\psig}{\psi(T)} -
    \bracket{\psi(T)}{\psig}\bracket{\psig}{\delta \psi(T)} \\
    & \qquad - |\bracket{\psig}{\delta \psi(T)}|^2
  \end{split} \\
  &= -
  2\mathrm{Re}\left\{\bracket{\psi(T)}{\psig}\bracket{\psig}{\delta
      \psi(T)}\right\} - |\bracket{\psig}{\delta \psi(T)}|^2.
\end{align}
The last term is $\mathcal{O}(\delta \psi(T)^2)$, and since $\delta
\psi(T)$ is small we set these terms to zero. Hence, we have finally
\begin{equation}
  \delta J_{1,\psi} = - 2\mathrm{Re}\left\{\bracket{\psi(T)}{\psig}\bracket{\psig}{\delta
      \psi(T)}\right\}.
\end{equation}
Repeating this treatment for $\delta J_{2,\psi}$, we have
\begin{align}
  \begin{split}
    \nonumber \delta J_{2,\psi} &= 2\mathrm{Re} \int_0^T
    \bra{\chi(t)}(\partial_t + \tfrac{i}{\hbar}\hat{H})\ket{\psi(t) + \delta \psi(t)}
    \ud t \\ & \qquad - 2\mathrm{Re} \int_0^T \bra{\chi(t)}(\partial_t +
    \tfrac{i}{\hbar}\hat{H})\ket{\psi(t)}\ud t
  \end{split}
  \\
  \nonumber &= 2\mathrm{Re}\int_0^T \bra{\chi(t)}(\partial_t +
  \tfrac{i}{\hbar}\hat{H})\ket{\delta \psi(t)}\ud t\\
  \begin{split}
    &= 2\mathrm{Re}\Bigl\{\bracket{\chi(T)}{\delta \psi(T)} -
      \bracket{\chi(0)}{\delta \psi(0)}\Bigr. \\ & \qquad - \Bigl.\int_0^T
      \left(\bra{\chi(t)}\tfrac{i}{\hbar}\hat{H} - \bra{\partial_t
          \chi(t)}\right)\ket{\delta \psi(t)}\ud t\Bigr\}.
  \end{split}
\end{align}
Noting that the initial state is fixed, we must have $\delta \psi(0) =
0$. Thus setting $\delta J_\psi = 0$, we obtain the two equations
\begin{align}\label{eq:cond_1}
  \bracket{\psi(T)}{\psig}\bracket{\psig}{\delta
    \psi(T)} + \bracket{\chi(T)}{\delta \psi(T)} &= 0, \\ \label{eq:cond_2}
  \left(\bra{\chi(t)}\tfrac{i}{\hbar}\hat{H} - \bra{\partial_t
      \chi(t)}\right)\ket{\delta \psi(t)} &= 0.
\end{align}
Since these must be valid for an arbitrary choice of $\ket{\delta
  \psi}$, we obtain from Eq.\eqref{eq:cond_1} the boundary condition
\begin{equation}
  \label{eq:chi_bc}
  \ket{\chi(T)} = \ket{\psig}\bracket{\psig}{\psi(T)},
\end{equation}
and from Eq.\eqref{eq:cond_2} the equation of motion
\begin{equation}
  \label{eq:chi_em}
  i \hbar \partial_t \ket{\chi(t)} = \hat{H}\ket{\chi(t)}.
\end{equation}

We now continue the derivation by finding the variation $\delta J_\chi$,
which results in the condition already given in
Eq.~\eqref{eq:schroedinger}, namely that $\ket{\psi}$ must obey the
Schr\"{o}dinger equation. The variation $\delta J_{\varepsilon_i}$
results in the condition
\begin{equation}
  \label{eq:cond_3}
  - \frac{2}{\lambda_i}\, \mathrm{Im}\left\{\bra{\chi(t)}\frac{1}{\hbar}\frac{\partial \hat{H}}{\partial
      \varepsilon_i}\ket{\psi(t)}\right\} = 0,
\end{equation}
where $\lambda_i$ can be used to suppress updates at $t=0$ and $t=T$. It can be clearly seen, that Eq.\eqref{eq:cond_3} cannot be solved directly for the controls
$\varepsilon_i$ since we have a system of split boundary conditions: $\ket{\psi(t)}$ is only specified at $t = 0$, and similarly $\ket{\chi(t)}$ only at $t = T$. Hence we require an iterative scheme which will
solve the equations self-consistently.

\paragraph{Deriving an iterative scheme}

The goal of any iterative method will be to reduce the objective $J$ at
each iteration while satisfying the constraints. Written mathematically,
we simply require that $J^{k+1} - J^{k} < 0$, where $J^{k}$ is the value
of the functional $J$ evaluated at the $k$th iteration of the
algorithm. We will also attach this notation to other objects to denote
which iteration of the algorithm we are referring to. Looking at
Eq.~\eqref{eq:cond_3} and taking into account our constraints, the
optimization algorithm presents itself as follows:
\begin{enumerate}
\item Make an initial guess for the control fields
  $\varepsilon_i(t)$.

\item At the $k$th iteration, propagate the initial state
  $\ket{\psi(0)}$ until $\ket{\psi^{k}(T)}$ and store it at each
  time step $t$.

\item Calculate $\ket{\chi^{k}(T)}$ from
  Eq.~\eqref{eq:chi_bc}. Our initial guess from step (1) should have been
  good enough that the final overlap of the wavefunction with the goal
  state is not zero (otherwise Eq.~\eqref{eq:chi_bc} would give us
  $\ket{\chi^{k}(T)} = 0$).

\item Propagate $\ket{\chi^{k}(T)}$ backward in time in accordance with
  Eq.~\eqref{eq:chi_em}. At each time step, calculate the new control at
  time $t$
  \begin{equation}
    \label{eq:grad_upd}
    \varepsilon_i^{k+1}(t) = \varepsilon_i^{k}(t) + \gamma \frac{2}{\lambda_i}\,
    \mathrm{Im}\left\{
      \bra{\chi^{k}(t)}\frac{1}{\hbar}\frac{\partial \hat{H}}{\partial
        \varepsilon_i}\ket{\psi^{k}(t)}
      \right\},
  \end{equation}
  which one can identify as a gradient-type algorithm, using
  Eq.~\eqref{eq:cond_3} as the gradient. The parameter $\gamma$ is
  determined by a line search to ensure our condition that $J^{k+1} -
  J^{k} < 0$.

\item Repeat steps (2) to (4) until the desired convergence has been
  achieved.
\end{enumerate}
This gradient-type method, while guaranteeing convergence, is rather
slow. A much faster method is what is known as the Krotov method in the
literature. Here, the modified procedure is as follows:
\begin{enumerate}
\item Make an initial guess for the control fields
  $\varepsilon_i(t)$.

\item Propagate the initial state $\ket{\psi(0)}$ until $\ket{\psi(T)}$.

\item At the $k$th iteration, calculate $\ket{\chi^k(T)}$ from
  Eq.~\eqref{eq:chi_bc} (again taking care that the final state overlap
  with the goal state is non-zero).

\item Propagate $\ket{\chi^{k}(T)}$ backward in time in accordance with
  Eq.~\eqref{eq:chi_em} to obtain $\ket{\chi(0)}$, storing it at each
  time step.

\item Start again with $\ket{\psi(0)}$, and calculate the new control at
  time $t$
  \begin{equation}
    \label{eq:krotov_upd}
    \varepsilon_i^{k+1}(t) = \varepsilon_i^{k}(t) + \frac{2}{\lambda_i}\, \mathrm{Im}\left\{\bra{\chi^{k}(t)}\frac{1}{\hbar}\frac{\partial \hat{H}}{\partial
        \varepsilon_i^k}\ket{\psi^{k+1}(t)}\right\}.
  \end{equation}
Use these new controls to propagate $\ket{\psi^{k+1}(0)}$ to obtain
$\ket{\psi^{k+1}(T)}$.

\item Repeat steps (3) to (5) until the desired convergence has been
  achieved.
\end{enumerate}
This new method looks very similar to the gradient method, except that
now we see that we must not use the `old' $\ket{\psi^k(t)}$ in the
update, but the `new' $\ket{\psi^{k+1}(t)}$. This is achieved by
immediately propagating the current $\ket{\psi^{k+1}(t)}$ with the newly
updated pulse, and not the old one. To make this explicit, take at $t =
0$, $\ket{\psi^{k+1}(0)} = \ket{\psi(0)}$. We use this to calculate the
first update to the controls $\varepsilon_i^{k+1}(0)$ from
Eq.~\eqref{eq:krotov_upd}. We use these controls to find
$\ket{\psi^{k+1}(\Delta t)}$, where $\Delta t$ is one time-step of our
simulation. We then obtain the next update $\varepsilon_i^{k+1}(\Delta
t)$ by again using Eq.~\eqref{eq:krotov_upd} where we use the old
$\ket{\chi^k(\Delta t)}$ that we had saved from the previous step. In
other words, $\ket{\chi(t)}$ is always propagated with the old controls,
and $\ket{\psi(t)}$ is always propagated with the new controls.

For a full treatment of this method in the literature, see
\citet{Sklarz2002}. For our purposes, we simply note that the method is
proven to be convergent, meaning $J^{k+1} - J^{k} < 0$, and that it has
a fast convergence when compared to many other optimization algorithms,
notably the gradient method \cite{Somloi1993}.
\subsection{Application}
\begin{figure}
\includegraphics[width=0.5\textwidth,angle=0]{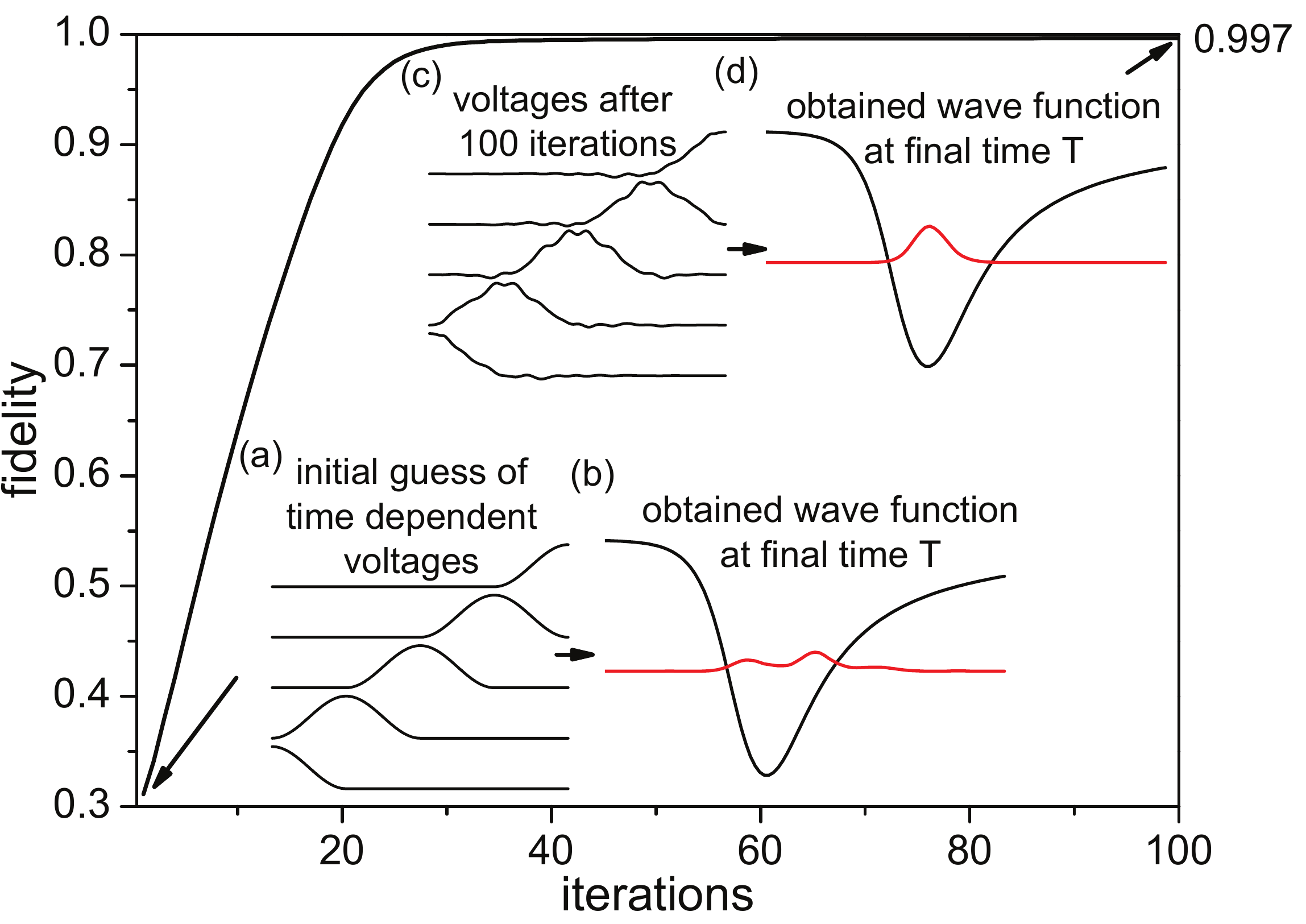}
\caption{(Color online). Fidelity increase after iterative optimization steps. (a) Initial guess of the time-dependent voltages applied to the electrodes. (b) Resulting wavefunction at the final time $T$. The fidelity is only 0.3. (c) Voltage configurations obtained after 100 iterative calls of the Krotov optimal control method. (d) The final wavefunction obtained at time $T$ with optimized voltages agrees well with the target ground state wavefunction. The fidelity is increased to $0.997$.  } \label{fig:fidel_trans}
\end{figure}
We now show how the wavefunction can be controlled via tailored time-dependent external potentials. In ion-trap experiments quantum control of the wavefunction via individual electrodes is difficult, as these are several orders of magnitude larger than the size of the wavefunction. However, matching length scales occur for cold atoms trapped in optical lattice potentials \citep{Calarco:2004} or magnetic micro trap. We nonetheless focus on the ion trap system and present this as a generic example for quantum control where we would like to transport the ion from one place to another without exciting higher motional states. The transport is performed by applying time-dependent voltages $u_i(t)$ to the electrodes generating a total potential
\begin{equation}
 \Phi(x,u_1(t),u_2(t), ...,u_n(t))=\sum_{i=1}^{n} \Phi_i(x)u_i(t).
 \end{equation}
The Hamiltonian of this system is
\begin{equation}
H_0(x,u_1(t),...,u_n(t))=-\frac{\hbar^2}{2m}\frac{d^2}{dx^2}+\Phi(x,u_1(t),...,u_n(t))\label{eq:h0}.
\end{equation}
As a target wavefunction  $\ket{\psig}$, we choose a harmonic oscillator ground state wavefunction centered at the target position. We thus have to maximize the wavefunction overlap  $|\bracket{\psig}{\psi(T)}|^2$, where $\ket{\psi(T)}$ is the wavefunction at the final time $T$ obtained by application of the time-dependent voltages. This is exactly the fidelity functional of Eq.~\eqref{eq:fidelity} and the initial condition for the Langrange multiplier of Eq.~\eqref{eq:chi_bc}. The update Eq.~\eqref{eq:krotov_upd} for the control parameters $u_i^k(t)$ from iteration step $k$ to $k+1$ has the form
\begin{equation}
u_i^{k+1}(t)=u_i^{k}(t)+\frac{2}{\lambda_i}\bra{\chi^k(t)}\frac{1}{\hbar}\Phi_i(x)\ket{\psi^{k+1}(t)}.
\end{equation}
Starting with a sinusoidal-shaped initial guess for the time-dependent voltage configuration $u_i^0(t)$ (see Fig.~\ref{fig:fidel_trans}(a)) the wavefunction at the final time $T$ is excited to the first excited state, leading to a wavefunction overlap of 0.3 as seen in Fig.~\ref{fig:fidel_trans}(b). After 100 steps of optimization the wavefunction overlap has been iteratively increased to 0.997 (Fig.~\ref{fig:fidel_trans}) and the motional ground state has been preserved (see Fig.~\ref{fig:fidel_trans}(d)). The optimized time-dependent voltages can be seen in Fig.~\ref{fig:fidel_trans}(c). The full source code of the optimal control algorithm to perform the presented optimization is contained in the \textit{octtool} package.

\section{Improving quantum gates -- OCT of a Unitary transformation }
Up to now we have only considered the motional degrees of freedom of trapped ions. The internal electronic degrees of freedom serve as information storage for the qubit and are manipulated by laser or microwave radiation. They are subject to quantum dynamics where a certain degree of sophistication can arise due to the presence of interference effects. Joint manipulation of the internal and external degrees of freedom is therefore a promising playground for the application of OCT theory along with the quantum dynamical simulation methods presented in chapter~\ref{sec:TDSE}. In the following we present the Cirac-Zoller controlled-not gate as a case study. We first explain the gate mechanism, then we use the OCT algorithm to derive a laser pulse sequence for the proper realization of the quantum gate.
\label{sec:OCTgate}
\subsection{The Cirac-Zoller controlled-not gate}
Atomic qubits are suitable carriers of quantum information because one can exploit the internal electronic degrees of freedom for realizing a near-perfect two level system \citep{CohenTannoudji:2004}, where coherent transitions between the states can be driven by laser radiation. Suitable long lived states are found to be sublevels of the electronic ground state connected via stimulated Raman transitions \citep{Poschinger:2009} or metastable electronic states excited on narrow dipole forbidden transitions \citep{SchmidtKaler:2003}. For neutral atoms the additional possibility of employing Rydberg states exists \citep{Gaetan:2009}. In the following we will explain the \citet{Cirac:1995} controlled-not (\textsc{cnot}) gate as realized by \citet{SchmidtKaler:2003b}. To understand each single step of the gate operation we first have to become acquainted with the light-ion interaction \citep{Leibfried:2003}. In the following $\left|\downarrow \right>$ and $\left|\uparrow \right>$ denote the qubit states with energies 0 and $\hbar \omega_0$, respectively. The full Hamiltonian of the system is $H=H_0+H_a+H_L$, where $H_0$ is given by Eq.~\eqref{eq:h0} with a constant harmonic trap potential $\Phi(x)=m\omega_{\textrm{tr}}^2x^2/2$, $H_a=\left|\uparrow\right> \left< \uparrow \right| \hbar \omega_0$ describes the energy of the internal electronic excitation.  $H_L$ describes the interaction between light and atom \citep{Sasura:2002,James:1998}\footnote{For a dipole transition $H_L$ can be written as $H_L^d=-e \vec{r}\cdot \vec{E}$,
and similarly for a quadrupole transition $H_L^q=-\frac{e}{2} \sum_{i,j} r_i r_j \frac{\partial E_j}{\partial x_i}$, with $\vec{E}=\vec{E_0} \textrm{cos}(\vec{k} \cdot \vec{x} - \omega_L t-\phi)$,
where $\vec{r}$ denotes the relative position of the valence electron with respect to the nucleus and $\vec{x}$ is the position of the ion. The frequency of the laser is given by $\omega_L$, with the optical phase $\phi$. We obtain the matrix elements of $H_L$ by means of the identity operator $\mathrm{1\!l}=\left|\downarrow\right> \left< \downarrow \right|+\left|\uparrow\right> \left< \uparrow \right|$ such that $H_L=\mathrm{1\!l} H_L \mathrm{1\!l}$. All diagonal elements vanish and the Hamiltonian can be written as in Eq.~\eqref{eq:hl}, where we have expressed the cosine by exponentials.}:
\begin{equation}
H_L=\frac{\hbar \Omega}{2} \left(\left|\uparrow\right> \left< \downarrow \right|+\left|\downarrow\right> \left< \uparrow \right|\right) \left( e^{i(\vec{k}\cdot \vec{r} - \omega_L t-\phi)}+\mathrm{h.c.}\right)\label{eq:hl},
\end{equation}
where $\Omega$ is the Rabi frequency of the transition between the qubit states. In this form, the Hamiltonian contains terms oscillating at the laser frequency. For optical frequencies we would need rather small time steps for an accurate numerical simulation. To avoid this, we change to the interaction picture
\begin{eqnarray}
\ket{\psi}_I&=&e^{i H_a t / \hbar} \ket{\psi},\quad \textrm{and}\\
H_I&=&e^{i H_a t / \hbar} H e^{-i H_a t / \hbar},
\end{eqnarray}
where $\ket{\psi}$ is the state of the motional and internal degrees of freedom in the Schr\"odinger picture. $H_I$ can be expanded by using the Baker-Campbell-Hausdorff formula\footnote{$e^A B e^{-A}=\sum_0^\infty [A,B]_m 1/m!$ with $[A,B]_m=[X,[X,Y]_{m-1}]$, the commutator $[A,B]=AB-BA$ and $[A,B]_0=B$.}
\begin{equation}
H_I=H_0+\frac{\hbar \Omega}{2}\left( e^{i\omega_0 t} \left|\uparrow\right> \left< \downarrow \right|+\mathrm{h.c.} \right) \left( e^{i ( k x - \omega_L t - \phi)} +\mathrm{h.c.} \right).
\end{equation}
 If we additionally make the rotating wave approximation, i.e. neglect fast oscillating terms at the frequency $\omega_L+\omega_0$, we obtain
\begin{equation}
H_I=H_0+\frac{\hbar \Omega}{2}\left[\left|
\uparrow \right> \left< \downarrow \right| e^{i ( k x - \delta t - \phi)}  + \mathrm{h.c.}\right]\label{eq:hi},
\end{equation}
with the laser detuning $\delta=\omega_{\textrm{L}}-\omega_{0}$. Now we can do the numerics with much larger time steps. The term proportional to $ \left|
\uparrow \right> \left< \downarrow \right|e^{i k x}$ is responsible for absorption processes: it changes the $\left|\downarrow\right>$ state into the $\left|\uparrow\right>$ state and displaces the motional state in momentum space by the photon recoil $\hbar k$. The Hermitian conjugate term proportional to  $ \left|
 \downarrow \right> \left< \uparrow \right|e^{-i k x}$ is responsible for stimulated emission processes from the $\ket{\uparrow}$ state back to the $\ket{\downarrow}$ ground state, where a photon is emitted back into the laser field displacing the motional state in momentum space by $-\hbar k$.

If the laser frequency $\omega_{\textrm{L}}$ is tuned to the
atomic resonance, such that $\delta=0$, then the interaction
describes simple Rabi oscillations between the qubit states with frequency $\Omega$. No net momentum is transferred as absorption processes and stimulated emission contribute equally. Thus, we can use this interaction for direct control of the internal states without changing the motional state. If the laser is irradiated on the ion during the time $t$ such that $\Omega t = \pi/2$ (referred to as a $\pi/2$ pulse), then superposition states are created\footnote{We have subsequently omitted all normalization factors $1/\sqrt{2}$ as they do not change the physical interpretation.}: $\ket{\downarrow} \rightarrow \ket{\downarrow} + \ket{\uparrow}$ and  $\ket{\uparrow} \rightarrow -\ket{\downarrow} + \ket{\uparrow}$. These states evolve as $\ket{\downarrow} + e^{- i \omega_0 t}\ket{\uparrow}$. If we now apply a second $\pi/2$ pulse in phase with the  oscillating superposition, we obtain the states $\ket{\downarrow} + \ket{\uparrow}\rightarrow \ket{\uparrow}$  and $  -\ket{\downarrow} + \ket{\uparrow} \rightarrow - \ket{\downarrow}$. If the optical phase is shifted by $\pi$ such that the laser field and the superposition are oscillating out of phase, we reverse the action of the first $\pi/2$ pulse: $\ket{\downarrow} + \ket{\uparrow} \rightarrow \ket{\downarrow} $  and $  -\ket{\downarrow} + \ket{\uparrow} \rightarrow  \ket{\uparrow}$. Hence, we obtain orthogonal results depending on the phase relation between the laser and the superposition state, which is the basic principle of Ramsey spectroscopy. If the laser frequency is kept perfectly resonant and phase stability is maintained, then one can detect externally induced phase flips of the superposition state during the waiting time $T$ by mapping the phase to the two states $\ket{\downarrow}$ and $\ket{\uparrow}$. Starting from the ground state, application of resonant $\pi/2$ pulses continuously cycles through the series of states
\begin{eqnarray}
\ket{\downarrow}\xrightarrow{\pi/2} \ket{\downarrow}+\ket{\uparrow}\xrightarrow{\pi/2}\ket{\uparrow}\nonumber\\\xrightarrow{\pi/2} -\ket{\downarrow}+\ket{\uparrow}\xrightarrow{\pi/2} -\ket{\downarrow}.\label{eq:rabi}
\end{eqnarray}
We can see that a concatenation of four $\pi/2$ pulses (a $2\pi$ pulse) takes the system back to the ground state, however with a phase flip of $\pi$, which is due to the fundamental $4\pi$ rotational symmetry of spin-1/2 systems\footnote{A global phase does not have any physical significance and can always be absorbed into the definition of the states $\ket{\psi}'\rightarrow e^{i\phi}\ket{\psi}$. The absolute laser phase does not matter before the first laser pulse starts, of importance is the relative phase of the superposition (imprinted by the first laser pulse) and a subsequent laser pulse.}.

When the laser frequency is tuned below the atomic resonance
by the vibrational trap frequency $\omega_{\textrm{tr}}$,
i.e. the laser is red detuned by $\delta=- \omega_{\textrm{tr}}$, we drive red-sideband
transitions between the states $\ket{\downarrow,n}$ and $\ket{\uparrow,n-1}$
with reduced Rabi frequency $\Omega \eta \sqrt{n}$. Rabi oscillations are obtained in a similar manner as in Eq.~\eqref{eq:rabi} by replacing $\ket{\downarrow}\rightarrow |\downarrow,n\rangle$ and $\ket{\uparrow}\rightarrow \ket{\uparrow,n-1}$. $n=0,1,2,\dots$ denotes the harmonic oscillator eigenstates, and $\eta=x_0\cdot k$ is the Lamb-Dicke parameter, where $x_0$ is the size of the ground state wavefunction. This parameter sets the coupling strength between the laser radiation and the atomic motion. Atomic excitation on the red sideband is accompanied by the lowering of the harmonic oscillator energy by one vibrational quantum.

For a blue laser-detuning with $\delta=\omega_{\textrm{tr}}$, the blue sideband interaction is realized. On the blue sideband transitions between the states $\ket{\downarrow,n}$ and
$\ket{\uparrow,n+1}$ are driven with Rabi frequency $\Omega \eta \sqrt{n+1}$. When applying a $\pi$-pulse to the $\ket{\downarrow,0}$ state we excite one motional quantum and obtain the state $\ket{\uparrow,1}$. On the other hand, if applied to the $\ket{\uparrow,0}$ state, no motional quantum can be excited thus this state does not couple to the blue sideband. A $\pi$ pulse on the blue sideband transition can therefore be used to map quantum information back and forth between the internal state of a specific ion and the motional state of an ion chain if a collective vibrational mode is driven. This operation is referred in the following as a \textsc{swap} operation.

\begin{figure}
\includegraphics[width=0.5\textwidth,angle=0]{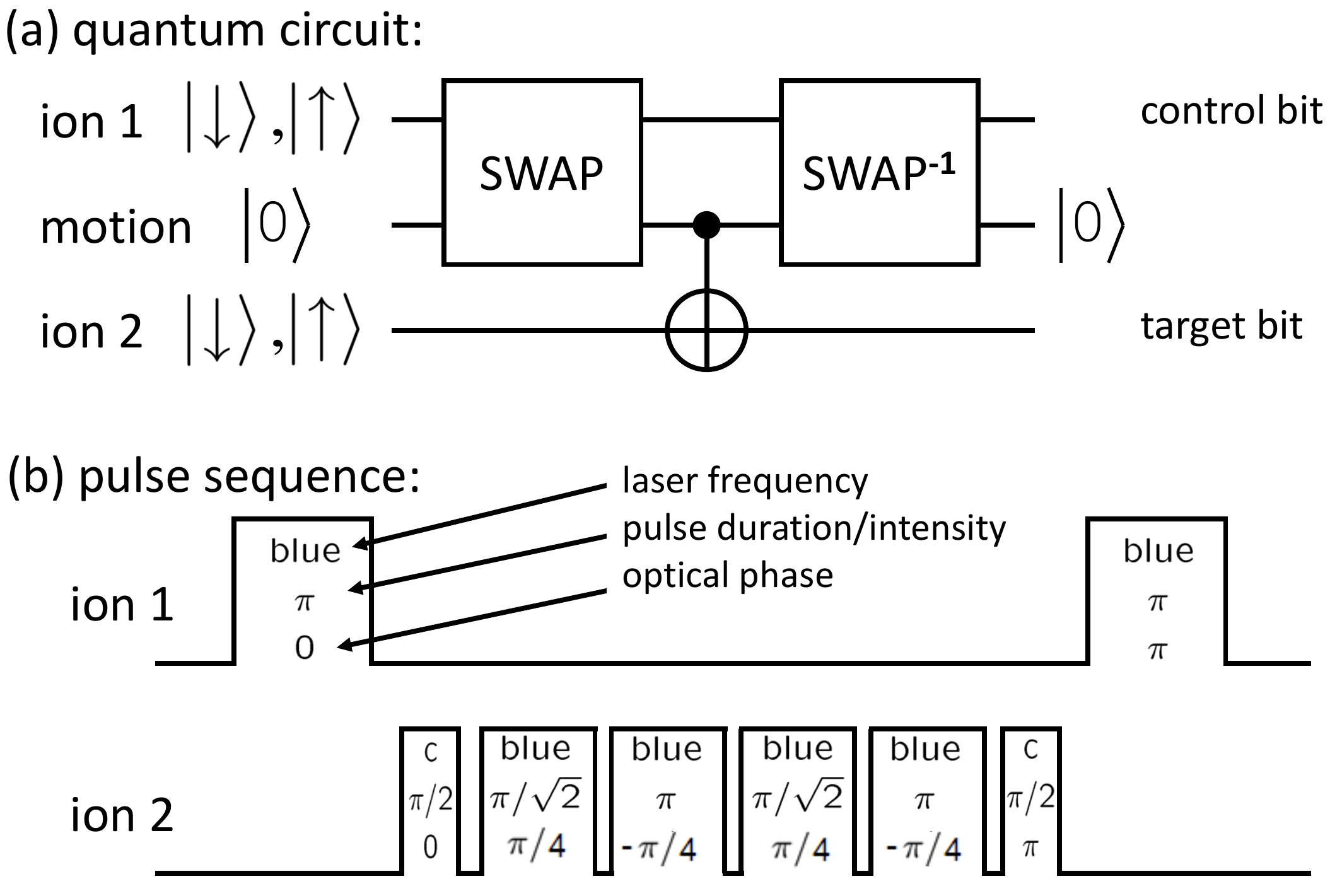}
\caption{(a) Quantum circuit for a \textsc{cnot} gate between two ions realized by a \textsc{swap} operation on the control ion, a \textsc{cnot} gate between the motional state and the target ion and a \textsc{swap}$^{-1}$ operation. (b) Composite pulse sequence to realize the total \textsc{cnot} gate between two ions \citep{SchmidtKaler:2003b,Schmidt-Kaler:2003}.} \label{fig:cnot}
\end{figure}

Now we have all the tools at hand to put the quantum gate together. In the following one ion will be referred to as control ion whose internal state is denoted by $\ket{\,\cdot\,}_c$, and a second target ion with internal state $\ket{\,\cdot\,}_t$. We want to flip the state of the target ion conditionally to the state of the control ion, realizing the \textsc{cnot} truth table:
\begin{eqnarray} \nonumber
\ket{\downarrow}_\textrm{c}\ket{\downarrow}_\textrm{t}&\rightarrow&\ket{\downarrow}_\textrm{c}\ket{\uparrow}_\textrm{t},\\ \nonumber
\ket{\downarrow}_\textrm{c}\ket{\uparrow}_\textrm{t}&\rightarrow&\ket{\downarrow}_\textrm{c}\ket{\downarrow}_\textrm{t},\\ \nonumber
\ket{\uparrow}_\textrm{c}\ket{\downarrow}_\textrm{t}&\rightarrow&\ket{\uparrow}_\textrm{c}\ket{\downarrow}_\textrm{t},\\
\ket{\uparrow}_\textrm{c}\ket{\uparrow}_\textrm{t}&\rightarrow&\ket{\uparrow}_\textrm{c}\ket{\uparrow}_\textrm{t}.
\end{eqnarray}
This gate is performed by the following steps: First the state of the control ion is mapped on a collective vibrational mode of the two ions by means of a \textsc{swap} operation. The remaining task is to perform a \textsc{cnot} gate between the vibrational mode and the target ion and finally perform the \textsc{swap}$^{-1}$ operation to restore the state of the control ion (see Fig.~\ref{fig:cnot}(a)). The \textsc{cnot} gate between the motional mode and the internal state of the target ion corresponds to the truth table
\begin{eqnarray} \nonumber
\ket{\downarrow,0}_\textrm{t}&\rightarrow&\ket{\uparrow,0}_\textrm{t},\\ \nonumber
\ket{\uparrow,0}_\textrm{t}&\rightarrow&\ket{\downarrow,0}_\textrm{t},\\ \nonumber
\ket{\downarrow,1}_\textrm{t}&\rightarrow&\ket{\downarrow,1}_\textrm{t},\\
\ket{\uparrow,1}_\textrm{t}&\rightarrow&\ket{\uparrow,1}_\textrm{t},\label{eq:CNOTmot}
\end{eqnarray}
where the motional state now acts as the control. The key element of this operation is a \textit{controlled phase} gate between these two degrees of freedom, which corresponds to the truth table
\begin{eqnarray} \nonumber
\ket{\downarrow,0}_\textrm{t}&\rightarrow&-\ket{\downarrow,0}_\textrm{t},\\ \nonumber
\ket{\uparrow,0}_\textrm{t}&\rightarrow&\ket{\uparrow,0}_\textrm{t},\\ \nonumber
\ket{\downarrow,1}_\textrm{t}&\rightarrow&-\ket{\downarrow,1}_\textrm{t},\\
\ket{\uparrow,1}_\textrm{t}&\rightarrow&-\ket{\uparrow,1}_\textrm{t}.\label{eq:CNOTphase}
\end{eqnarray}
As explained above, mapping of a superposition phase to the states is accomplished by means of resonant $\pi/2$ pulses. If the controlled phase is sandwiched between two such pulses, then the internal state of the target ion is given by the conditional phase from the phase gate operation. This realizes the \textsc{cnot} gate of Eq.~\eqref{eq:CNOTmot}.

The phase gate itself is performed by exploiting on the one hand the fact that the blue sideband does not couple to the $\ket{\uparrow,0}$ state and on the other hand that a $2\pi$ pulse flips the phase of any given state by $\pi$ as can be seen from Eq.~\eqref{eq:rabi}. Therefore a $2\pi$ pulse on the blue side band changes the phase for the states $\ket{\downarrow,0}$, $\ket{\uparrow,1}$ and $\ket{\downarrow,1}$, whereas $\ket{\uparrow,0}$ is left unchanged. This would result in the conditional phase gate of Eq.~\eqref{eq:CNOTphase} such that the whole \textsc{cnot} gate sequence is complete.

Additional complications arise due to the fact that the blue
sideband Rabi frequency on the $\ket{\downarrow,1}\rightarrow
\ket{\uparrow,2}$ state is larger than the one on the
$\ket{\downarrow,0}\rightarrow \ket{\uparrow,1}$ transition by a
factor of $\sqrt{2}$. The problem was resolved by a theoretical proposal by \citet{Childs:2000} and realized experimentally by \citet{SchmidtKaler:2003b,Schmidt-Kaler:2003} through the application of a composite pulse sequence
of blue sideband pulses with different durations and phases as
seen in Fig.~\ref{fig:cnot}(b). In the next section
we will demonstrate how such sequences can be automatically
obtained by application of quantum optimal control techniques.

\subsection{Krotov algorithm on unitary transformations}
Now we show how the optimal control algorithm from Sec.~\ref{sec:OCT} finds a control sequence which realizes the controlled phase gate. The input for the optimal control algorithm is the system dynamics governed by the TDSE and the Hamiltonian $H_I$ from Eq.~\eqref{eq:hi}. The subject to be controlled is the unitary transform of Eq.~\eqref{eq:CNOTphase}. The state of the system is represented by two distinct wavefunctions in position space for the states $\ket{\uparrow}$ and $\ket{\downarrow}$. We now perform the optimal control algorithm for a unitary transformation along the lines of \citet{Palao2002}. Instead of one initial and one target state, we now have four initial states $\ket{\psi_s(0)}$ and four target states $\ket{\psig_s}$  ($s=1...4$) corresponding to the states in Eq.~\eqref{eq:CNOTphase}. Additionally, we have to change the fidelity objective of Eq.~\eqref{eq:fidelity} to a phase sensitive one with
\begin{equation}
\tilde{F}[\psi] \equiv  \frac{1}{8} \left(\sum_{s=1}^4 \bracket{\psig_s}{\psi_s(T)}\right)+\frac{1}{2},
\end{equation}
such that we have $J_1[\psi] \equiv -\tilde{F}[\psi]$. The constraint is now that the TDSE is to be fulfilled for all four states, thus we introduce four Lagrange multipliers $\ket{\chi_s(t)}$. Eq.~\eqref{eq:tdse_const} is now changed into
\begin{align}
    J_2[\varepsilon_i, \psi, \chi] &\equiv  2 \sum_{s=1}^4 \mathrm{Re} \int_0^T \bra{\chi_s(t)}(\partial_t +
      \tfrac{i}{\hbar}\hat{H_I})\ket{\psi_s(t)}\ud t
\end{align}
From these equations we derive the initial condition for the Lagrange multipliers $\bra{\chi_s(T)}=\bra{\psig_s}$ and the update equation
 \begin{equation}
    \label{eq:krotov_upd2}
    \varepsilon_i^{k+1}(t) = \varepsilon_i^{k}(t) + \frac{2}{\lambda_i}\, \sum_{s=1}^4\mathrm{Im}\left\{\bra{\chi_s^{k}(t)}\frac{i}{\hbar}\frac{\partial \hat{H_I}}{\partial
        \varepsilon_i^k}\ket{\psi_s^{k+1}(t)}\right\}.
  \end{equation}

\subsection{Application}
\begin{figure}
\includegraphics[width=0.5\textwidth,angle=0]{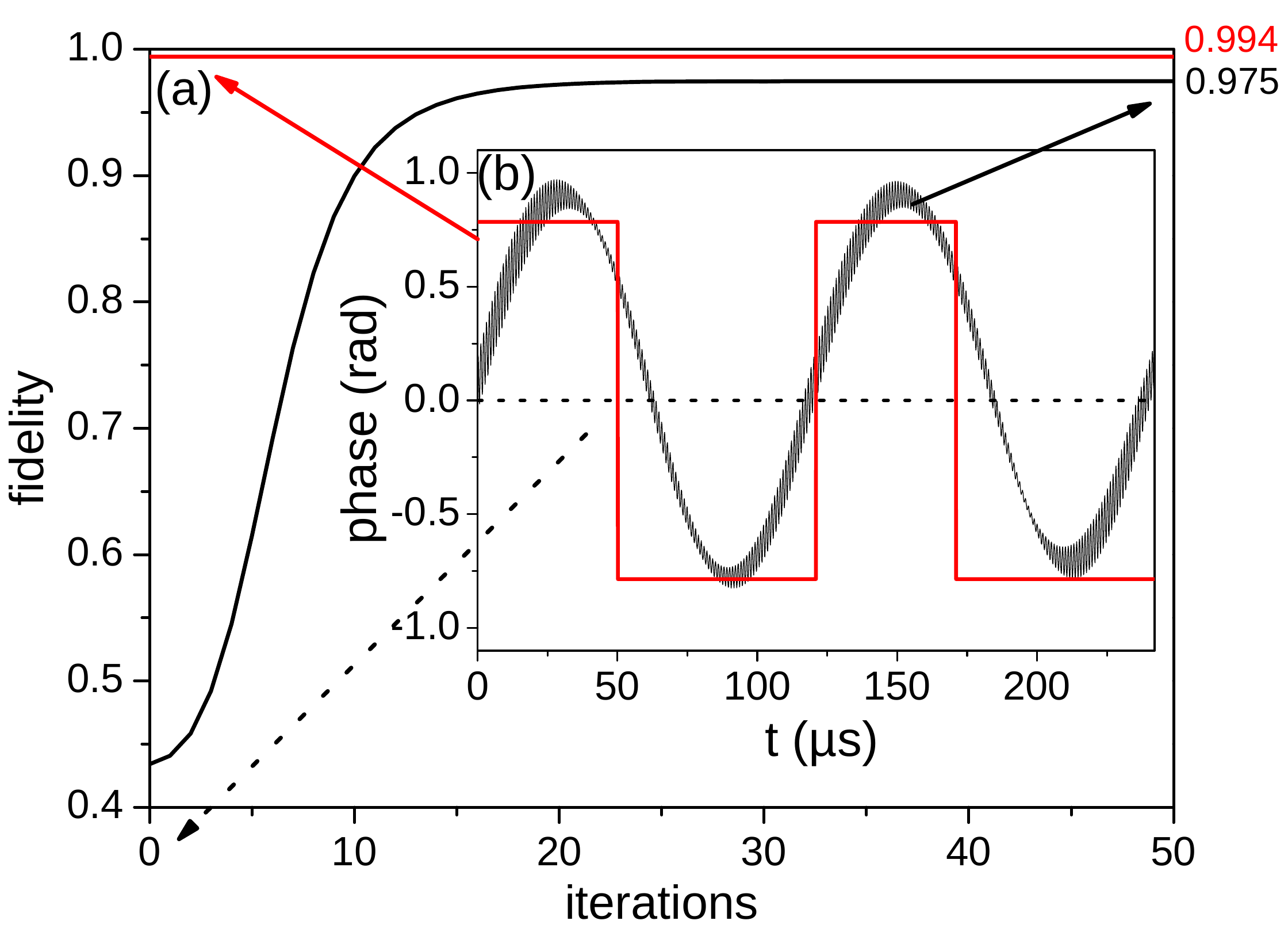}
\caption{(Color online) (a) Increase in fidelity from 0.43 to 0.975 after 50 iterations (black). The composite pulse sequence realizes a fidelity of 0.994 (red). (b) Time-dependent phase applied on the blue sideband. The dotted curve shows the inital guess $\phi(t)=0$ and the black curve shows the result obtained after 50 iterations of the Krotov algorithm. Note the similarity in the obtained phases concerning amplitude and period when compared to the composite pulse sequence as used in \cite{SchmidtKaler:2003b,Schmidt-Kaler:2003} (red curve). } \label{fig:gate}
\end{figure}

We now compare the four pulses on the blue sideband (see Fig.~\ref{fig:cnot}(b)) to the pulse found by the optimal control algorithm. For our gate optimization problem we need only one control parameter which is the phase $\phi$ of the laser on the blue sideband $\varepsilon_1(t)\equiv \phi(t)$ with the initial guess $\phi(t)=0$. Fig~\ref{fig:gate}(a) shows the increase of the fidelity from 0.43 to 0.975 after 50 iterations. The composite pulse sequence used in \cite{SchmidtKaler:2003b,Schmidt-Kaler:2003} achieves a fidelity of 0.994. In both cases, the deviation from unity is caused by off-resonant excitation on the carrier transition. It is quite remarkable that the Krotov algorithm finds a time-dependent phase $\phi(t)$ with similar amplitude and period as the composite pulse sequence (see Fig.~\ref{fig:gate}(b)). We have illustrated the OCT method on an example of a quantum control problem where a solution is already known. For more complex control problems however, this might not be the case, such that OCT allows for tackling control problems where the vastness of Hilbert space and the complexity of quantum interference are hard to handle manually. All sources can be found in the \textit{octtool} package, where additionally a simulation of the gate operation on the wavefunction is visualized. Further application of optimal control techniques to quantum gates with ions can be found in the literature \citep{Zhu:2006,GarciaRipoll:2003}.

\section{Conclusion}
\label{sec:Conclusion}
\textit{Applicability to other qubit types:}
Currently, trapped ion quantum systems are leading experimental efforts in quantum information theory. But with the growing maturity of quantum information experiments with trapped neutral atoms or solid state systems, we stress that the methods presented here will be applicable also to these systems with minor modifications. In this section we briefly elucidate how far each of the methods presented is applicable to each type of qubit and indicate, where appropriate, the connections between them with relevant citations. The methods from Sec.~\ref{sec:ClassicalTrajectories} for the numerical calculation of particle trajectories are directly applicable to neutral atoms, where it might also be interesting to analyze their motional behavior in trap structures like magnetic microchip traps~\cite{Fortagh:2007}, which have grown greatly in complexity, essentially realizing labs on atom chips. The ion trap community has mimicked the success story of atom chips by the development of microstructured ion traps. For neutral atoms, the full quantum dynamical simulations are of an even higher importance than for ion trap systems, which is due to the fact that neutral atoms are generally more weakly confined and are therefore much more sensitive to anharmonicities of the trap potentials. Quantum dynamical simulations have been used in conjunction with the optimal control method from Sec.~\ref{sec:OCT} to investigate the perspectives for transport and splitting operations in magnetic microtraps and optical lattices~\cite{Calarco:2004,Treutlein:2006, DeChiara:2008, Hohenester:2007,Riedel:2010}. The optimal control method might turn out to be of essential importance for the realization of robust high-fidelity quantum gates in artificial atom systems like Josephson junction-based qubits or impurity-based qubits in solid state host materials~\cite{Kane:1998, Neumann:2008}, where the level scheme, the coupling to external control fields and the decoherence mechanisms are generally more complicated than for trapped ion qubits~\cite{Spoerl:2007,Montangero:2007}. In general, the ability to calculate the quantum dynamics for any kind of engineered quantum system is of fundamental importance, and the methods presented in Sec.~\ref{sec:TDSE} can be directly adapted to any given Hamiltonian. The methods for the precise and efficient calculation of electrostatic fields from Sec.~\ref{sec:field} are also of interest for the optimization of electrode geometries for quantum dot qubits based on a two dimensional electron gas~\cite{Vandersypen:2006}. Furthermore, the Laplace equation is of a similar mathematical structure as the Helmholtz equation for ac electromagnetic fields, such that these fields may be calculated in miniaturized microwave traps for neutral atoms or microstructured Josephson devices based on adapted fast multipole methods\cite{Gumerov:2004}.

\textit{Summary:}
We have presented the whole process of ion trap quantum computing starting from trap design, trapping, and transport of ions, and ending with laser-ion interactions and the simulation and optimization of the Cirac-Zoller \textsc{cnot} gate starting from basic principles. The explanation of the physics is complemented by a detailed description of the numerical methods needed to perform precise simulations, which are carefully selected such that both precision and efficiency are maintained. Additionally, the numerical methods are presented in a general manner, such that they may be applied to solve physical problems outside of the particular focus of this paper. The source code of all methods together with all needed libraries packed into a single installation file can be downloaded from \texttt{http://kilian-singer.de/ent} for both Linux and Windows operating systems. For fast testing of the routines we have bundled them with the C++ scripting library \textsc{root}\footnote{Website at \texttt{http://root.cern.ch}.}. By making these libraries accessible to the public we want to inspire students to experiment, and provide researchers with a foundation to perform more sophisticated simulations.
\bibliographystyle{apsrmp}

\section{Acknowledgements}
This work has been supported by the European Commission projects EMALI and FASTQUAST, the Bulgarian NSF grants VU-F-205/06,
VU-I-301/07, D002-90/08, and the Elite program of the Landesstiftung Baden-W\"{u}rttemberg. The authors
thank David Tannor, Ronnie Kosloff and Christiane Koch for useful discussions and Rainer Reichle for his contributions at an early stage of the project.

\bibliography{rmp2008}
\end{document}